# Emergent Transport Properties of Molecular Motor Ensemble Affected by Single Motor Mutations

Shreyas Bhaban*[1], Donatello Materassi[2], Mingang Li[3], Thomas Hays[3], and Murti Salapaka[1]

[1] *Department of Electrical Engineering, University of Minnesota-Twin Cities, Minneapolis MN 55455, USA*
[2] *Department of Electrical Engineering and Computer Science, University of Tennessee-Knoxville, Knoxville, TN 37996, USA*
[2] *Department of Genetics, Cell Biology, and Development, University of Minnesota-Twin Cities, Minneapolis MN 55455, USA*

## Abstract

Intracellular transport is an essential function in eucaryotic cells, facilitated by motor proteins - proteins converting chemical energy into kinetic energy. It is understood that motor proteins work in teams enabling unidirectional and bidirectional transport of intracellular cargo over long distances. Disruptions of the underlying transport mechanisms, often caused by mutations that alter single motor characteristics, are known to cause neurodegenerative diseases. For example, phosphorylation of kinesin motor domain at the serine residue is implicated in Huntington's disease, with a recent study of phosphorylated and phosphomimetic serine residues indicating lowered single motor stalling forces. In this article we report the effects of mutations of this nature on transport properties of cargo carried by multiple wild-type and mutant motors. Results indicate that mutants with altered stall forces might determine the average velocity and run-length even when they are outnumbered by wild type motors in the ensemble. It is shown that mutants gain a competitive advantage and lead to an increase in the expected run-length when the load on the cargo is in the vicinity of the mutant's stalling force or a multiple of its stalling force.

A separate contribution of this article is the development of a semi-analytic method to analyze transport of cargo by multiple motors of multiple types. The technique determines transition rates between various relative configurations of motors carrying the cargo using the transition rates between various absolute configurations. This enables an exact computation of biologically relevant quantities like average velocity and run-length. It can also be used to introduce alterations of various single motor parameters to model a mutation and to deduce effects of such alterations on the transport of a common cargo by multiple motors. Our method is easily implementable and we provide a software package for general use.

*bhaba001@umn.edu

# Introduction

Motor proteins- kinesin, dynein and myosins- are nanoscale machines that are the main effectors of intracellular transport. They play a critical role in the growth and sustenance of healthy cells by enabling a transport of intracellular cargo over networks of microtubules [1]. Disruption of the functions performed by these molecular motors is linked to neurodegenerative diseases such as Huntington's, Parkinson's and Alzheimer's Disease [2, 3], muscular disorders such as heart disease, uterine complications and high blood pressure. The mechano-chemical behavior of a single motor moving along the microtubule substrate to transport a cargo is relatively well understood [4]. However *in vivo* it is known that multiple motors work in teams to transport a common cargo [5]; how multiple motor-proteins coordinate to transport a common cargo is not well understood [5, 6, 7]. Many studies employ a probabilistic description of the behavior of a single motor protein to construct models that describe how multiple motors transport a common cargo [5, 8, 9, 10]. In [11], Gross and coworkers employed Monte-Carlo simulation studies built on a model of a single kinesin in to explore how multiple identical kinesin motors might interact to transport a cargo against a hindering load force. Their work indicates, counter intuitively, the existence of a form of strain-gating, where the motors of an ensemble share loads unequally enabling cargo transport over longer distances. Xu and coworkers examined the effects of ATP concentration on the transport of cargo carried by single and two motors in [12]. At decreased levels of ATP concentration, the velocities of cargoes transported by single and two motor proteins decreases. Coincidentally at decreased ATP concentration there was an an appreciable increase in the run-length of cargoes transported by two motors, while no such effect was seen in the case of transport by one motor. The authors proposed that the increased run-length observed in the presence of two motors results from the lowered dissociation of each motor from the microtubule at decreased ATP concentrations and the increased probability that the cargo stayed bound to the microtubule. Studies such as [10, 13] using probabilistic models of single motors have also predicted in that an ensemble of kinesin motors is a robust system and the robustness increases under high loads [14].

The study of cargo transport by a heterogeneous ensemble of motor proteins composed of both wild type and mutant motors is important to inform our understanding of how mutant motors impact intracellular transport and lead to an onser of diseases. Recent studies have implied that alterations in the kinesin-1 motor domain may have a role in impaired axonal transport. Phosphorylation of a mammal kinesin motor domain by kinase c-Jun N-terminal kinase-3 (JNK3) at a conserved serine residue (Ser-176 in A and C isoforms and Ser-175 in B isoform) is implicated in Huntington's disease [15]. However the mechanisms affected by Ser-175 phosphorylation are not well understood. An experimental study by Selvin and coworkers in [16] reported that a negative charge at Ser-175, acquired through mutation or phosphorylation, leads to a lower stall force and decreased velocity under external loads of 1 pN or more, while leaving the ATPase, microtubule-binding affnity and processivity unchanged.

Using a semi-analytical method, we reveal surprising emergent transport behaviors arising when a cargo is transported by multiple motor proteins, some of which are mutated and some are not. In particular we analyze the impact of Ser-175 kinesin mutation such as those reported in [16] on cargo transport in the presence of wild type motors. The detailed investigation made possible by our method leads us to hypothesize that under certain conditions, proteins *moor* the cargo to the microtubule and prevent it from being lost. While these *mooring proteins* do not contribute to the motion of the cargo, they enhance the probability of attachment of other cargo-bound motor proteins to the microtubule that subsequently contributes to an increase in average cargo displacement. The activation of mooring mechanism depends on a number of external factors such as load force and ATP concentration. However, it is also determined by intrinsic properties of the motor protein such as the stall force of the individual motors. Remarkably, mutant motors that have stalling forces matched to the external load force can act dominantly and determine emergent transport properties such as longer run-length, even when they are outnumbered in the ensemble by wild-type motors. Such mechanisms could potentially point to how diseased states emerge and progress coincident with the accumulation of the mutant motor species.

A separate contribution of the article is a semi-analytical method for determining the probability distribution of various configurations of a cargo carried by multiple number and types of motor-protein. A detailed experimental study of the various modalities of transport by multiple motor proteins (homologous or otherwise) requires significant instrumental resolution than what is needed to investigate single motor behavior. As a consequence, observing the transport dynamics of multiple motors is experimentally challenging [17, 18]. It is further compounded by the combinatorial complexity introduced by the multitude of scenarios possible when many motors and motor types participate in transport. Such challenges motivate the use of analytical and computational tools. The mean-field approach in [8, 10] makes use of simplifying assumptions, such as equal load sharing among all motors, to achieve analytical results, thereby sacrificing significant detail for computational benefits. The Monte Carlo approach in [11], provides better fidelity where complex models can be employed; however, the accuracy of results depend on the number of iterations and on the rarity of the events that occur. Unlike the Monte-Carlo simulations or any currently implemented simulation method, our Master Equation based method [19] analytically solves for the probability distributions of all possible scenarios at any time point. The methodology is uniquely powerful and enables the calculation of various biologically relevant quantities such as average velocity and run-length, for reasonably sized ensembles and with high accuracy while using lesser computational resources and time. The underlying concepts behind our methodology are motivated by earlier work reported in [13]. The key enabling concept is that of 'relative configuration' of motor proteins, determining the transition probabilities between relative configurations from the transition probabilities of the absolute configuration space and subsequently determining the biologically relevant quantities from the relative configuration space. The computation engine is implemented using MATLAB and can be used to simulate cargo transport by any two unidirectional species.

Our method provides a general platform to study the transport of cargo by multiple motors of two different types where each type of motor protein can be individually characterized by a probabilistic model describing its stepping, detachment and attachment rates. For this article, the technique has been utilized to introduce alterations of various parameters from the nominal ones to model a mutation of the serine residue and compute the effect of such a mutation on cargo carried by a mixture of wild-type and mutant motor-proteins. In summary, we developed a simulation engine to study the transport of cargo by multiple motor proteins with distinct properties that in concert can exhibit emergent transport behaviors.

# Results and Discussion

In this study, we investigate the impact of a previously reported kinesin mutation on the transport of a motor ensemble and its attached cargo. The mutation, located within the motor domain of the kinesin, mimics the phosphorylated state of Ser-175 [16]. The motor domain phosphomimetic mutation does not affect the ATPase, microtubule-binding affinity or processivity of the motor, but does reduce the stall force and velocity of kinesin under a load force. The *in vitro* phosphorylation of Ser-175 for a full length kinesin similarly reduces stall force and velocity of the motor. Both the mutant and in virto phosphorylated kinesin showed no other aberrant single motor behavior.

Here we use computational modelling to analyze the impact of the Ser-175 mutant kinesin on a heterogenous ensemble of motors and its transported cargo. The wild-type and mutant kinesin motors have a different stalling force; the wild type kinesin has a stall force of $F_s = 6pN$ and the mutated kinesin has a reduced stall force of $\bar{F}_s = 5.5pN$(see [16] for more detail). In our analysis we considered cargoes transported by the following motor ensembles : cargo with two wild-type (WW) motors , one wild-type and one mutant (WM) motors, two mutant (MM) motors , three wild-type (WWW) motors , two wild-type and one mutant (WWM) motors, one wild-type and two mutant (WMM)motors, and three mutant (MMM) motors.

## How External Loading Conditions Affect Behavior of Heterogenous Motor Ensemble

### Motor configuration when engaged to the microtubule approaches steady state

For cargoes carried by multiple motor proteins, transport occurs as each motor steps along the microtubule lattice, sequentially binding, translocating and releasing from the microtubule in a well defined mechanochemical cycle [5, 9, 20, 21]. Eventually, the condition where no motors are attached to the microtubule will be reached and the cargo will diffuse away from the microtubule. Thus it is evident that the only steady state is the condition when none of the motors of the ensemble carrying the cargo remain attached to the microtubule. Indeed, in all the cases considered where there are three or less motors on the cargo (i.e. $M \leq 3$) it is seen that for any specified initial probability distribution of motors engaged to the microtubule, the steady-state probability distribution has no motors attached to the microtubule i.e. cargo is eventually lost. Figure 1(a) shows the exactly computed probability distribution for the WM case, where with increasing time the probability of no motors attached (red curve) increases and eventually reaches one.

Given that the steady-state probability distribution is the trivial case of no motors attached, nothing much of value can be derived regarding the motor behavior from the trivial steady-state probability distribution. However the *conditional* probability distribution of the various configurations of motor proteins carrying a common cargo, *given that at least one motor is attached to the microtubule* (i.e. when the cargo is not lost), is of significant interest. Here it is *apriori* not evident whether the conditional probability distribution has a steady state.

Remarkably, the results show that under a constant load force on the cargo the conditional probability distribution of the various motor configurations reaches a steady state. The steady state distribution is independent of the initial probability distribution chosen (see for example, Figure 1(b) for the WM case). This indicates that the cargo-motor ensemble consisting of motors with different stalling forces is a robust system which, starting from any arbitrary initial condition, behaves according to a fixed distribution after some transient period. Furthermore, the existence of such a steady-state distribution, termed here as the *steady-state conditional pdf*, allows the computation of most of the variables of biological interest.

### Load forces on cargo affect conditional steady state distribution of motor ensemble

The conditional steady-state distribution is used to analyze transport of cargo with two attached motors and assumes that at least one motor is attached (Figure 2). Figure 2 shows the probabilities for the number of wild-type and mutated motor proteins that remain attached under varying load forces. The analysis indicates that when the cargo has two wild type motors attached (WW case), the probability that both motors remain attached to the microtubule peaks when load force is close to $12pN$. This equals twice the stalling force of a single wild type motor; indicating a relationship between number of motors attached (obtained using the steady state distribution) and applied load. Thus at these loads there exists a propensity to retain and not loose the cargo. Similarly, in the MM case (where the cargo is attached to two mutant motors) the probability that both motors are attached to the microtubule is maximum near $11pN$ which is twice the stalling force of the mutated motor. The WM ensemble displays lower peaks for the dual attachment of motors at both load forces of $11pN$ and $12pN$.

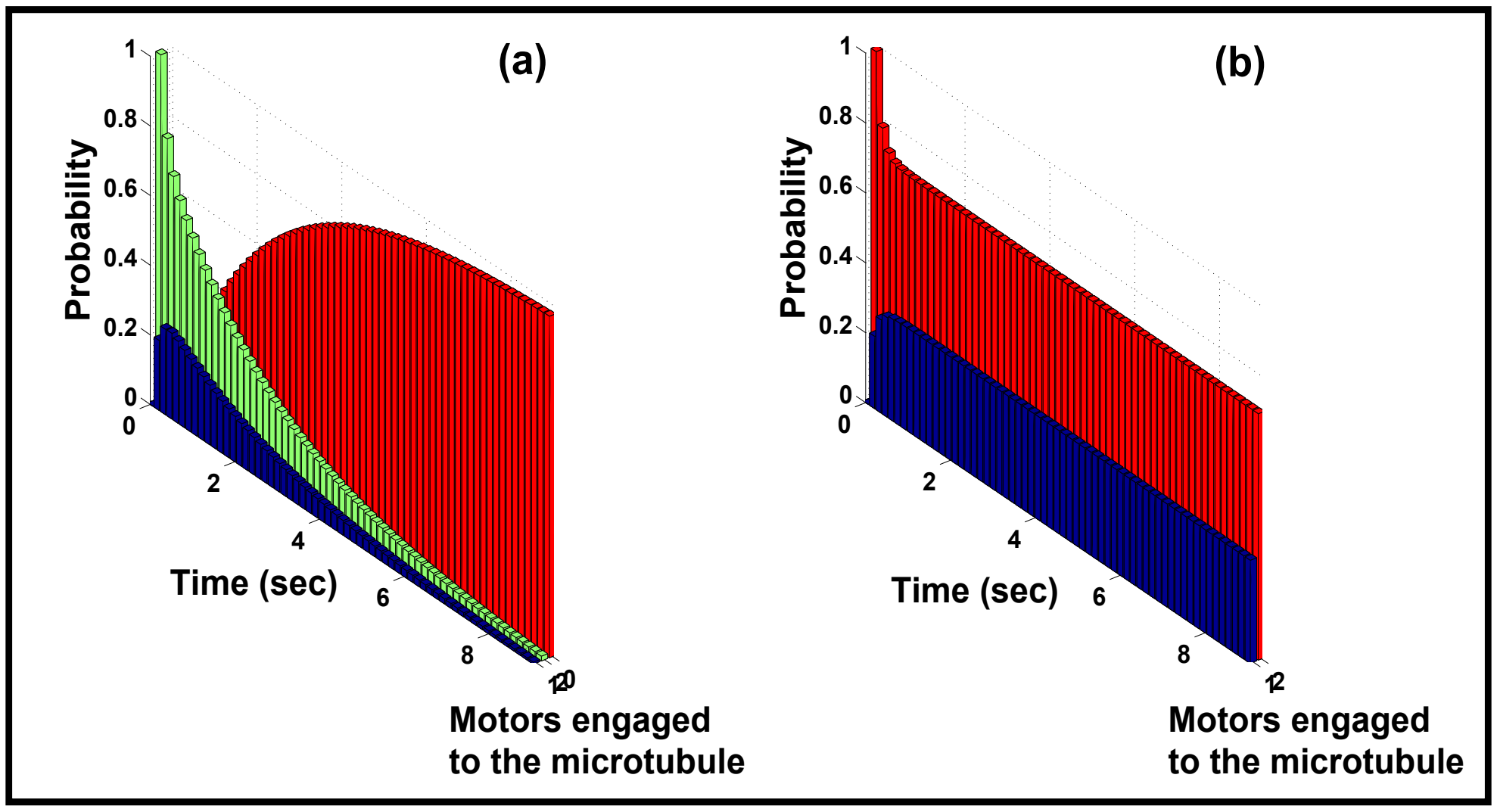

Figure 1: **Probability of motors being bound to the microtubule at time $t$ under load force of $1.2\ pN$ for the WM case.** In **(a)** the probability of one bound motor (blue) and two bound motors (green) approaches 0 as time $t \to \infty$. Probability of no bound motors (red) (i.e. the cargo dissociates from the microtubule) approaches 1 as $t \to \infty$. In **(b)** the probability of one bound motor (blue) and two bound motors (red) under the condition that cargo is not lost, reaches a steady state.

It is evident that for ensembles containing two motors, the conditional steady state distributions of the number of motors attached is not significantly affected until the load force on the cargo is close to twice the stalling force values of the wild -type or mutant motor ($2F_s = 12pN$ or $2\bar{F}_s = 11pN$). A similar observation is made for ensembles with three motors and for load force values close to twice and thrice the stalling force of a single motor or $2F_s$, $2\bar{F}_s$, $3F_s(= 18pN)$ and $3\bar{F}_s(= 16.5pN)$, pointing to a mechanism of equal sharing of loads in these load force ranges. Thus a cooperative mode between an ensemble of motors is encouraged when the load forces are close to the stall-force or multiples of the stall forces. For example, if a WW ensemble is subjected to load force close to $2F_s = 12\ pN$, unequal load sharing would subject one of the motors to a force greater than its stalling force $F_s = 6\ pN$. This will lead to its detachment and eventual transfer of the entire load onto the remaining motor. This, in turn, would gradually lead to the remaining single motor being detached from the microtubule, and the resultant loss of cargo attachment to the microtubule substrate. On the other hand, an equal load sharing will stall both the motors but ensure the retention of the cargo. A similar inference can be made for cargo transport by three motors and load forces approaching the values $F_s$, $\bar{F}_s$, $2F_s$ or $2\bar{F}_s$ (figure not shown).

Note that, it is possible when the load force is close to twice the stalling force, the cargo detaches from the microtubule with a high probability. In such a case, the previously mentioned conclusions reached about the probabilities by applying the condition that the cargo is not lost (i.e. at least one motor remains engaged to the microtubule) will not carry much significance. We then utilized the probabilities obtained without the conditioning to reach similar conclusions. Figures 3 and 4 show that the time taken for the cargo to be lost (i.e. time taken for the probability of detachment of the cargo to be close to one) is higher when the load force is close to but less than the appropriate stalling force or twice the appropriate stalling force value. The time taken for detachment is lesser for a load force of an intermediate value (compare with Figure 5). The figures show that if the load force is matched to an appropriate stall force or its multiples (for example, close to $2F_s$ for WW or $2\bar{F}_s$ for MM ensemble), the probability of detachment of the cargo from the microtubule becomes appreciable only after a long time has passed. Thus, the conclusions reached using the conditional distribution are also supported by results inferred from the entire configuration space with no restriction that the cargo remains attached. It implies that the study on the condition that 'the cargo is not lost' provides important insights.

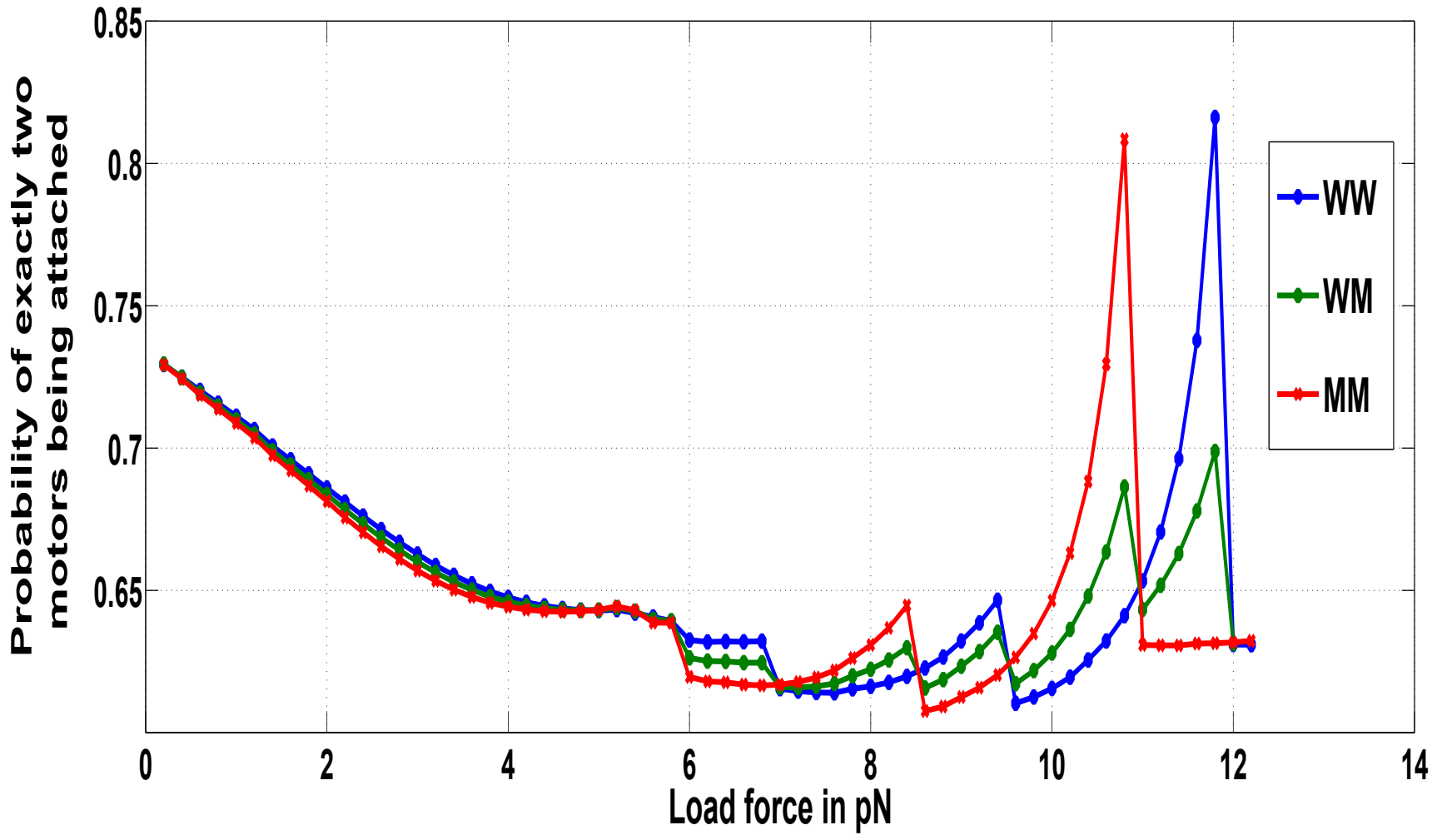

Figure 2: **Probability of exactly two motors attached to the microtubule for ensembles WW, WM and MM.**

In conclusion, load force on the cargo has a significant effect on the steady state distribution of motor ensembles and can provide an insight into how multiple motor ensembles function under varying loads. In particular, when a cargo is subjected to load forces that are multiples of the stalling force for individual motors, a cooperative behaviour in the form of equal load sharing is observed. This averts a scenario where unequal load sharing would subject some of the motors to loads greater than their stall forces, leading to their detachment and the eventual loss of cargo from the microtubule. Such cooperative behavior possibly helps retain the cargo. It is thus more probable that cargoes attached to a higher number of wild-type motors are retained under load forces close to multiples of the stalling force of a wild-type motor, making the presence of more mutants disadvantageous in these load force regimes.

## Hindering Load Force on the Cargo can Tune Multiple Motor Travel

The knowledge of the steady state conditional probability distributions allows one to compute biologically relevant quantities such as average velocity and run-length. The results obtained for ensembles

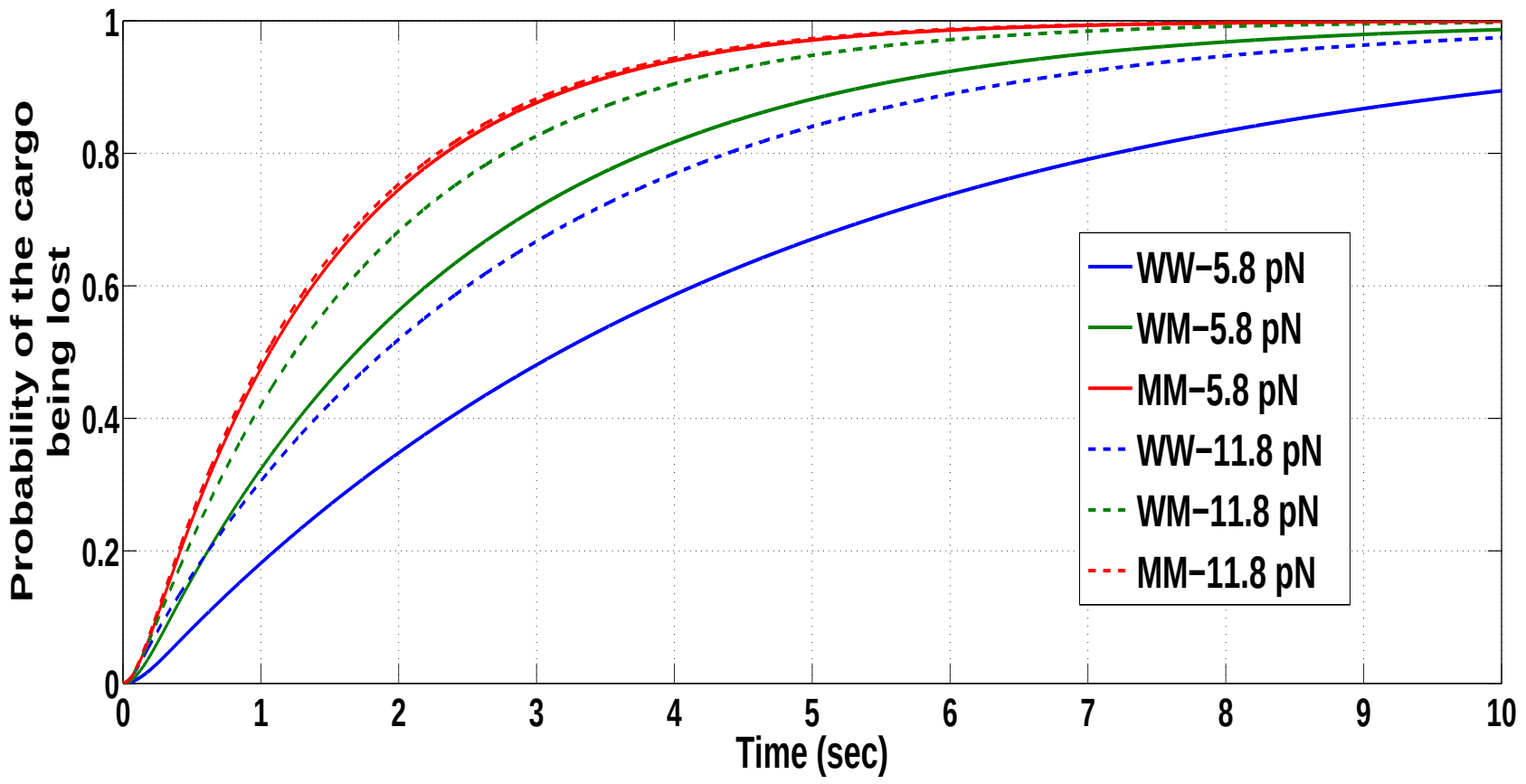

Figure 3: **Probability of cargo being lost for ensembles WW, WM and MM at load forces close to but less than $F_s$ and $2F_s$ ($F_s = 6\ pN$).**

containing two and three motors for varying load forces are reported.

### Average velocity typically reduces with increasing number of mutant motors, but varies as load forces approach multiples of stall forces

The effect of load force on the average velocity of a cargo is described in Figures 6 and 7 for a total of two and three attached motors, respectively. It is evident that for any given ensemble, average cargo velocity mostly reduces with increase in the hindering load force on the cargo. It is further evident from Figure 6 that for a given load force on the cargo, the average velocities are highest for the WW ensemble, followed by WM and the lowest for MM ensemble. For the same load force and number of motors in the ensemble, it follows that the average velocity decreases with increased participation of mutant motors (which have a lower stall force). A similar conclusion is reached when three motors are attached to the cargo (see Figure 7). Thus, for a fixed load force and total number of motors, average cargo velocity reduces with increasing number of mutant motors in the ensemble.

However, for some motors ensembles the trend of decreasing cargo velocity at elevated load forces is not observed. Figure 8 shows the average velocity for load forces between 10.4 $pN$ and 12.1 $pN$. In the case of WWW ensemble, as the load forces increase to 12 $pN$ (which is equal to $2F_s$) the average velocity *increases* with increasing load force. Our semi-analytic method allows for a fine analysis of such counter-intuitive observations where it is possible to extract the precise motor ensemble configurations that contribute to the observed effect.

Upon analysis it is observed that as the load force approaches the appropriate multiples of stalling force ($2F_s$ or $2\bar{F}_s$ in Figure 8), detachment of motors from the microtubule becomes less likely and the attached motors tend to cluster close together on the microtubule. This is similar to an observation made in [11, 13] where under high loads, the motors carrying a common cargo tend to remain clustered together while under low loads they tend to spread apart. Since the detachments are less likely, backward transitions of the cargo induced by single motor detachment are less frequent. Moreover, if a detachment were to occur, the magnitude of load induced backward transition of the cargo is less due to the clustering of motors under high loads. On average, this contributes to an increase in average velocity despite increasing load. This explains the increase in average velocities of the three motor ensembles at load forces approaching 12 $pN$.

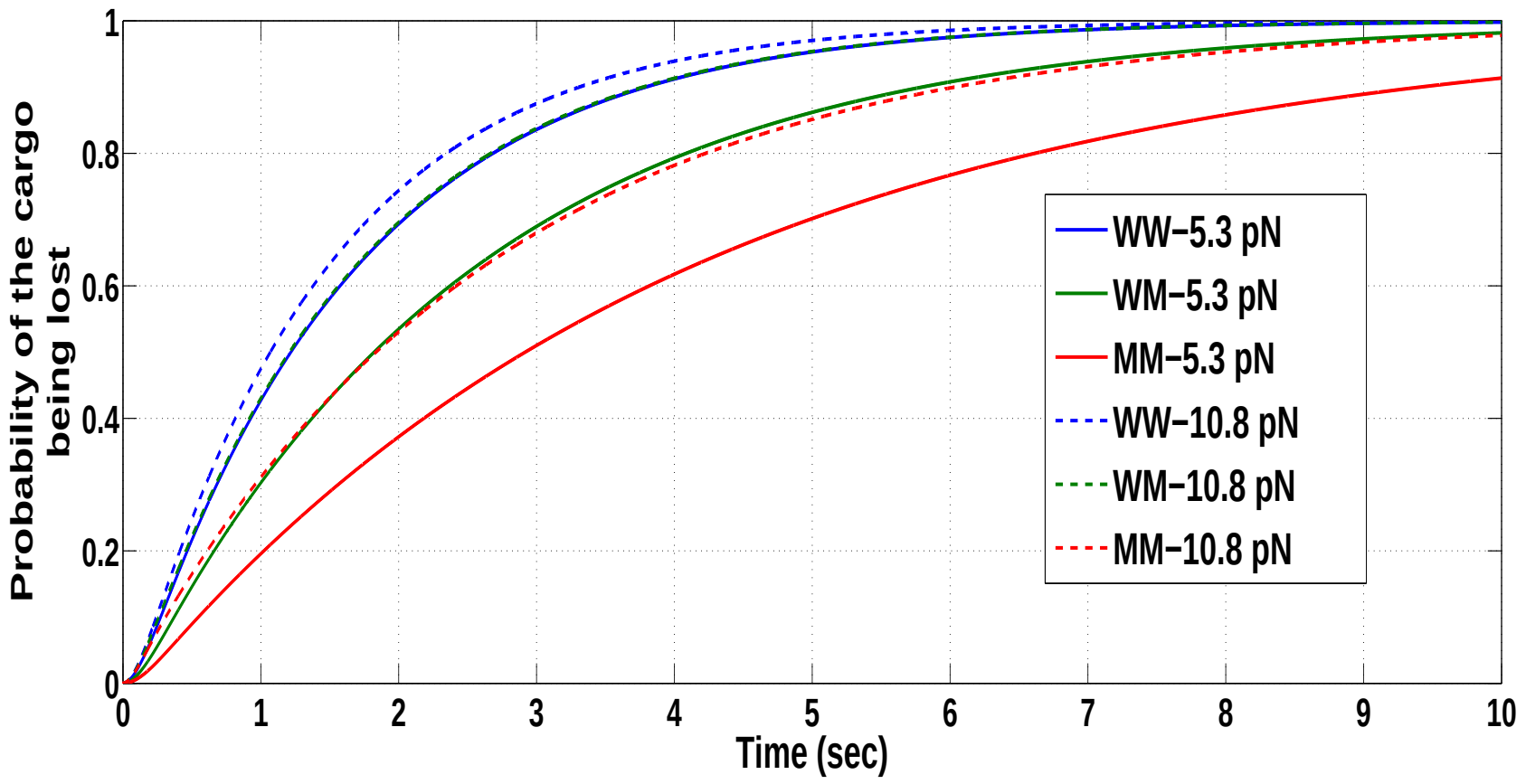

Figure 4: **Probability of cargo being lost for ensembles WW, WM and MM at load forces close to but less than $F_s$ and $2F_s$ ($F_s = 5.5\ pN$).**

Interestingly, the increase in average velocities at load forces approaching $2F_s = 12\ pN$ is more prominent in the WWW ensemble than in MMM ensemble, indicating that it is the detachment of the wild-type motors in the ensemble that becomes less unlikely. Thus, as the number of wild type motors increases, detachment events at load forces close to $2F_s = 12\ pN$ become less likely. Similarly, at load forces approaching $2\bar{F}_s = 11\ pN$ the average cargo velocity increases with increasing load force; with the increase becoming more prominent for ensembles having more numbers of mutant motors (i.e. higher for MMM ensemble than WWW ensemble).

Significantly, this observation suggests that by tuning the load forces on a cargo to values near the multiples of the stall force, one could counter a decreased velocity resulting from an increase in load force or the presence of a higher number of mutant motors.

An important feature of the semi-analytic method to be emphasized is the ease with which detection of rare events such as those described above are made possible. Monte Carlo based methodology, for example, would not only take significant computations to simulate rare events with a high degree of confidence, but would also be unable to determine the incidence causes of rare events. Thus our model allows the user to easily unearth the cause of a rare event.

In conclusion, load forces approaching multiples of the appropriate stall force for a cargo decrease the probability of detachment of associated motors from the microtubule. Since detachment of motors in a multiple motor ensemble can lead to backward motion of the cargo (because kinesin doesn't actively step backward i.e. towards minus end), a decrease in the detachment probability leads to higher forward motion of cargo on average, leading to an increase in average velocity even though the load increases to approach multiples of stalling forces. If load forces are close to multiples of $F_s(\text{or}\bar{F}_s)$ detachment of wild-type (or mutant) motors in the ensemble decreases and cargoes with higher wild-type (or mutant) motors demonstrate greater increase in average velocity.

### Cargo run-lengths impacted by heterogenity in the motor ensemble

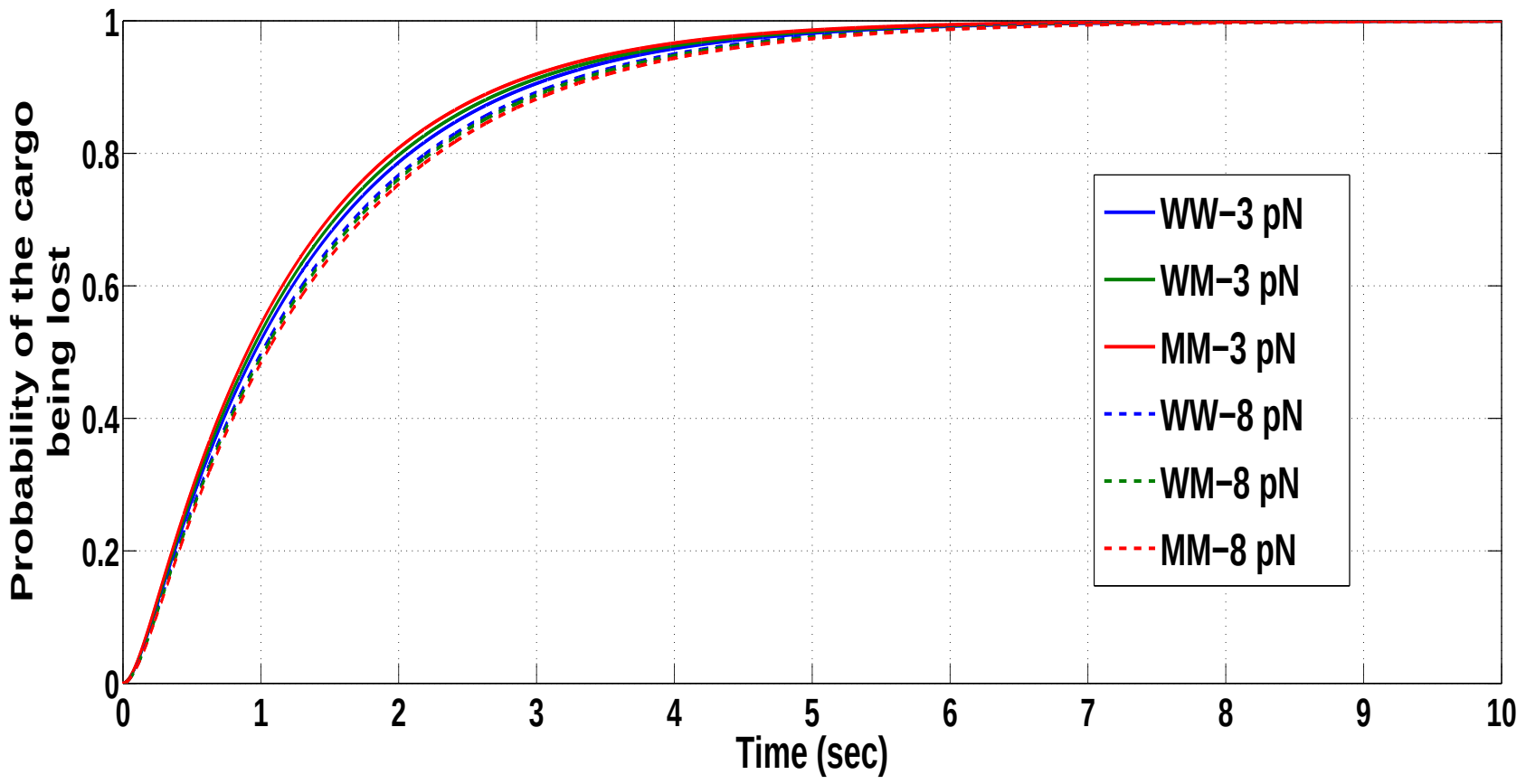

Figure 5: **Probability of cargo being lost for ensembles WW, WM and MM at load forces** 3 $pN$ **and** 8 $pN$

The dependence of run-length on the load force when the cargo has two and three motors attached to it is shown in Figures 9 and 10 respectively. A general trend is that the average run-length reduces with an increase in the antagonistic load force on the cargo. Moreover, for a fixed number of motors in an ensemble, at most values of load force, the run-length is *reduced* with an increasing number of mutant motors in the ensemble (Figure 9 inset(A) and Figure 10 inset(A)).

Surprisingly, under a load force approaching 6 $pN$ (Figure 9 inset(B)), there is a surge in the average distance traveled by the cargo, peaking close to 6 $pN$ and falling for larger load forces. The average run-length when load force is close to 6 $pN$ increases with the number of wild-type motors and decreases with the number of mutant motors (note that, at load force equal to 5.8 $pN$, ensemble WW travels $\approx$ 1000 $nm$, WM travels $\approx$ 500 $nm$ while MM travels $\approx$ 270 $nm$). Similar behavior is seen in Figure 10 insert(B). Note that 6 $pN$ is the stalling force $F_s$ of the wild-type motor.

Similarly, when the load force is close to 12 pN (Figure 10, inset C), the average run-length of a cargo is higher when the number of wild-type motors is greater that the number of mutant motors in the ensemble. Note that the overall run-length values for the three motor ensemble are lower here, since the load force on the cargo is higher (compare 12 $pN$, Figure 10 inset(C) versus 6 $pN$, Figure 10 inset(B)). Also note that 12 $pN$ is twice the wild-type motor stalling force $F_s$.

In Figure 9 inset(B), when the load force is close to 5.5 $pN$, a surge in the run-length is observed, peaking close to 5.5 $pN$ and falling for larger load forces. However, the order of increase in run-length seen here is opposite to that seen near 6 $pN$, i.e. the run-length is higher for larger number of mutant motors and is lower for a larger number of wild-type motors in the ensemble. Similar behavior is seen in Figure 10 inset(B). Note that 5.5 $pN$ is the stalling force $\bar{F}_s$ of the mutant motor.

In Figure 10 inset(C), for a load force close to 11 $pN$, the average run-length is higher when a larger number of mutant motors is present in the ensemble. Note that 11 $pN$ is twice the mutant motor stalling force $\bar{F}_s$.

To explain the reason behind such an observation, we examine the run-length observed for heterogeneous motor ensembles (e.g. WM). Using the conditional steady state probability distributions for the scenario where the load force is close to the wild-type stalling force, we determined the probability for the wild-type and mutant motor remaining attached to the microtubule (Figure 11). It is evident that the probability of the wild-type motor remaining engaged with the microtubule lattice is high when the load force is close to 6 $pN$. It can thus be inferred that under this condition, the cargo remains bound to the microtubule with a higher probability. In consequence, the detached motor has a greater opportuinity to reattach to the microtubule, overcoming the stalled condition of the leading motor and contributing to the increase in run-length. Thus the run-length under these conditions will be proportional to the number of wild-type motors, since it is the wild-type motors that will stall, tethering the cargo to the microtubule lattice and favoring its continued translocation. The peak in the run-length observed in Figure 7 inset(C) for a load force close to 12 $pN$ can similarly be explained. As the two leading wild-type motors stall and remain attached to the microtubule, the third motor will reattach, overcome the stalled condition and promote an extension of the run-length.

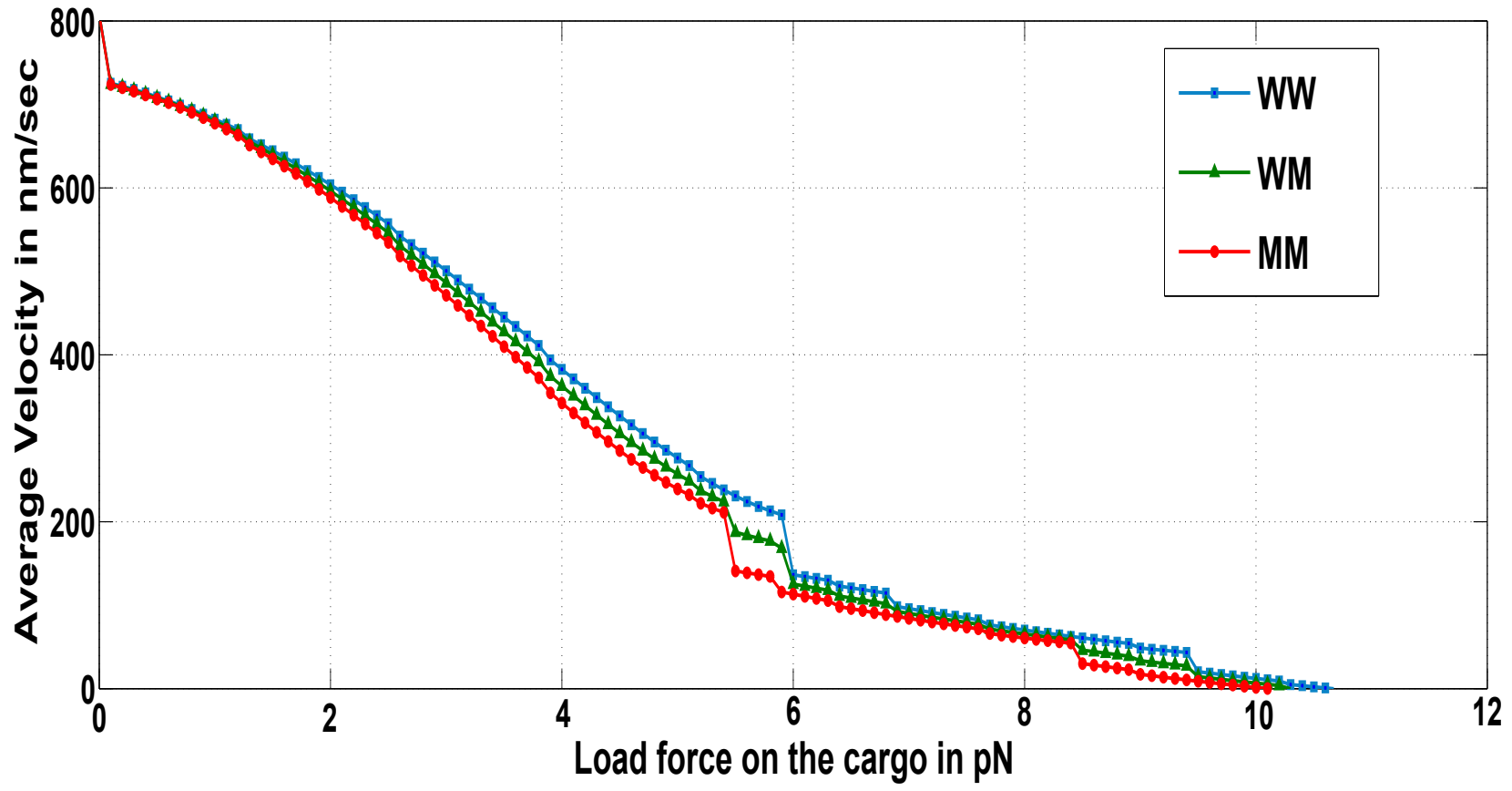

Figure 6: **Average velocity with varying load forces on the cargo for ensembles WW, WM and MM.**

The same load dependency of cargo run-lengths is also observed for mutant motors with reduced stall forces. At the reduced stall force, a mutant motor exhibits an increased probability for remaining attached to the microtubule lattice, preventing the cargo from diffusing away from the microtubule and increasing the runlength (Figure 11, see peak in green trace at $5.5pN$). The tethering by the stalled mutant motor increases the probability that the disengaged wild type motor will reattach, overcome the stalled mutant motor and extend the cargo runlength. The same effect also explains the positive correlation between run-length and number of mutant motors at load forces close to $\bar{F}_s$ and multiples of $\bar{F}_s$, as the probability of having mutant motors that are stalled increases making the loss of cargo more unlikely.

Significantly, in these load force ranges and for a fixed number of motors attached to the cargo, the cargo with more mutant motors attached will travel further than other cargoes with fewer or no mutant motors attached. Remarkably, the run-length of the cargo is predicted by the single motor characteristics, namely the stalling force, of the mutant motor.

### ATP Concentration can Tune Multiple Motor Travel

It is known that kinesin hydrolyzes one ATP per step [22]. Thus, the rate at which kinesin steps depends on its rate of ATP hydrolysis. It was shown experimentally in [12] that under no load conditions, average velocity for ensembles having one and two motors decreases with decreasing

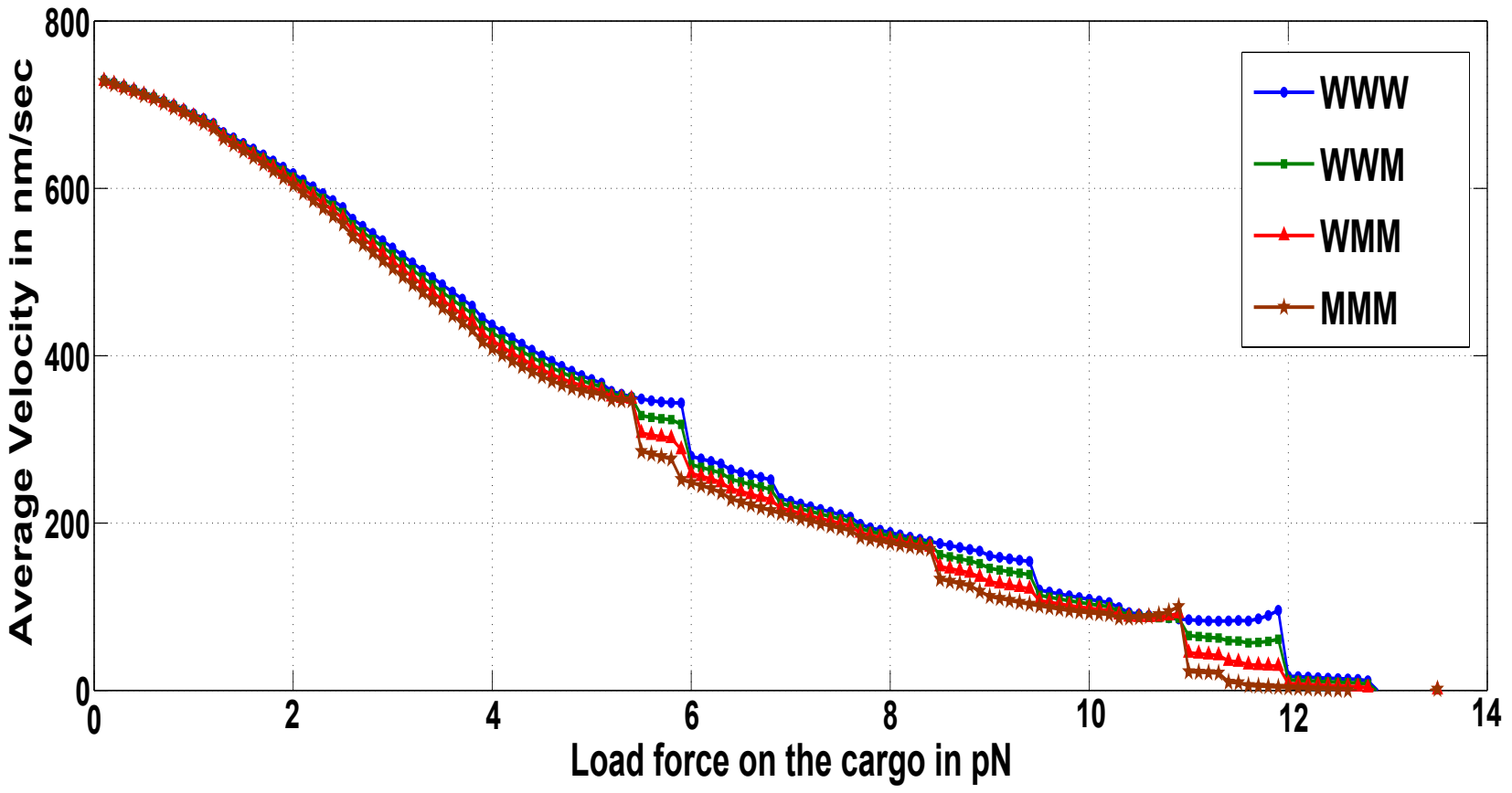

Figure 7: **Average velocity with varying load forces on the cargo for ensembles WWW, WWM, WMM and MMM.**

ATP concentration. It was also shown that run-length was not appreciably affected by changing ATP concentrations when the cargo had only one kinesin attached. However, when the cargo has two kinesin motors attached, run-length was demonstrated to have a strong negative correlation with single motor velocity (and single motor velocity was shown to have a positive correlation with ATP concentration). Our semi-analytic method was used to assess the conclusions reached in [12] and to further understand how the presence of mutant motors (with a different stalling force $\bar{F}_s$) can impact the ATP dependence of run length. The altering of this behavior due to the presence of more than two motors in the ensemble is also studied.

### Average run-length for multiple motor ensemble negatively correlates to ATP concentration when restricted to certain regimes

We first considered the affect of ATP concentration on run-length of a cargo containing only a single wild type motor (Figure 12). Our analysis shows that as ATP concentration increases, the distance traveled by the cargo also increases monotonically until ATP concentration saturates. When the cargo carried by a single wild-type motor is subjected to 0.2 $pN$ load force, the effect of ATP concentration on run-length is not substantial for concentrations above 10 $\mu M$ (611 $nm$ at 10 $\mu M$ ATP and 786 $nm$ at 2 $mM$ ATP). Thus run-length of a cargo with a single wild-type motor is not substantially affected by ATP concentration, agreeing with the trend observed in [12].

However, the effect of ATP concentration is more pronounced for ensembles containing more than one motor. For ensembles comprised of multiple motors, decreased ATP concentrations under certain conditions leads to increased average run-length of the cargo. For example, in both the WW (Fig 13) and WWW (Figure (S3Fig)) ensembles, we observe that with reducing ATP concentration, the run-length first increases, peaks at a certain value and then decreases. The average velocity of cargo transport will decrease with decreasing ATP concentration (see Figure(S1Fig) for WW and Figure(S2Fig) for WWW), indicating that the increased run-length is not due to increasing velocity but instead reflects the increased probability that the cargo remains attached to the microtubule for a longer time. As is evident from Figure 14 for WW ensemble, not only does the probability of at-tachment increase in the presence of two motors by comparison to one motor, but the probability that the motors remain attached to the microtubule also increases as the ATP concentration is decreased. A similar argument is provided for the experimental observations made by Xu and team in [12] for ensemble WW, that reported a negative correlation between single-motor velocities and run-length of cargo transport by ensemble WW.

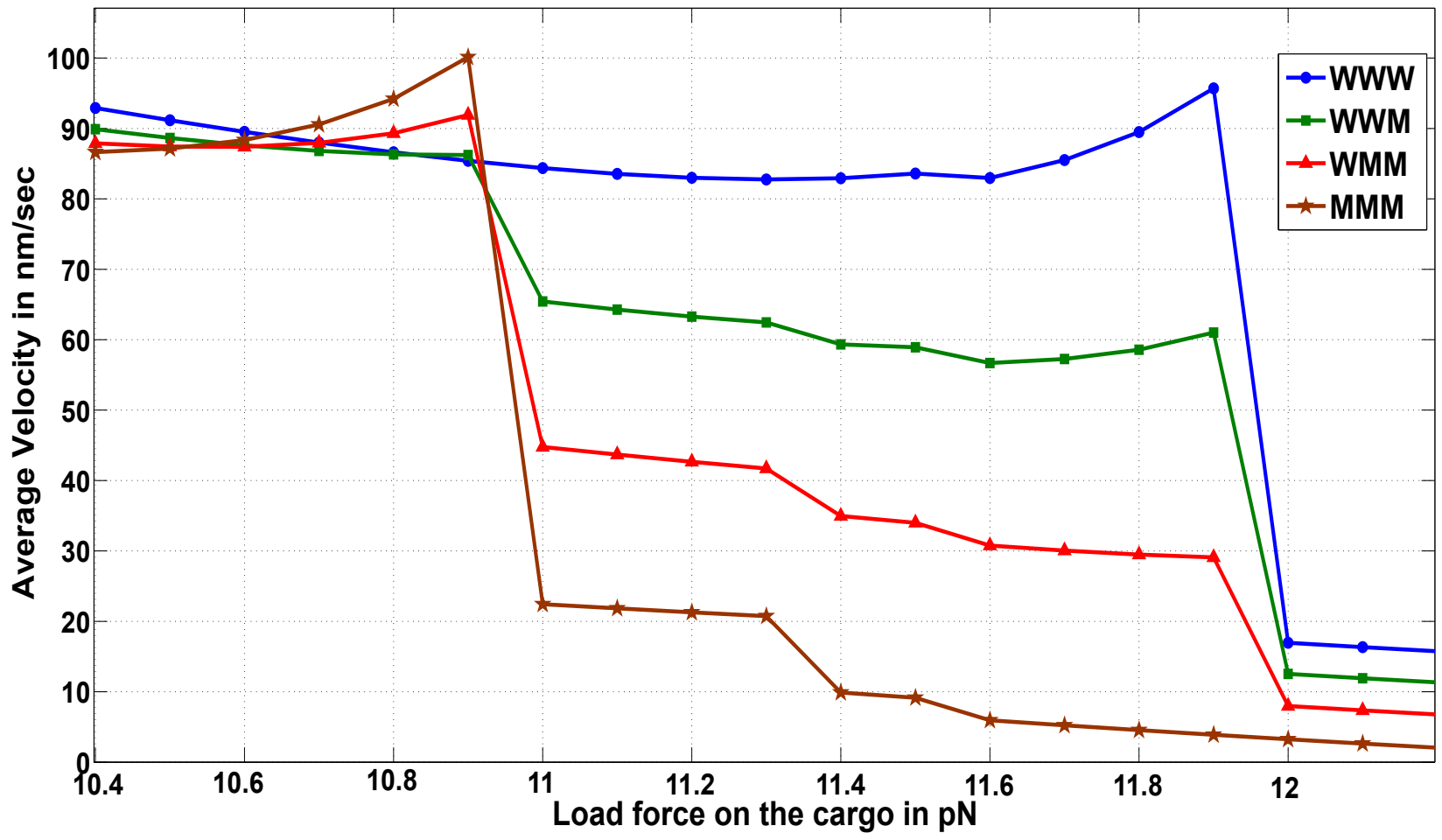

Figure 8: **Figure showing variation of average velocity with load force on the cargo for ensembles containing** $3$ **motors, for load forces between** $10.4\ pN$ **and** $12.1\ pN$**.**

Furthermore, the results also predict a negative correlation will persist until a certain minimum ATP concentration, below which the average run-length starts decreasing. To summarize, the run-length cannot continue to increase as ATP concentration falls, since at a certain threshold the low ATP concentrations will hinder ATP hydrolysis and the stepping of the motor domains along the microtubule lattice. At this limiting ATP concentration, regardless of how long the cargo remains attached, the motors will no longer be able to translocate along the microtubule lattice and runlength will decrease.

In conclusion, the ATP concentration in certain regimes negatively correlates with cargo run-length, as long as there are more than one motor attached to the cargo. Such behavior observed for a two motor ensemble agrees with results reported in [12].

### The heterogeneity of motors within an ensemble does not alter the effect of ATP concentration on run-length

To analyze the impact of ATP concentration on the behavior of ensembles with multiple motors, we characterized the average run-length and velocity for ensembles with two motors (Figure 13 and Figure(S1Fig)) and three motors (Figure(S3Fig) and Figure(S2Fig)).

Ensembles with mutant motors also exhibited a similar trend of increasing run-length as ATP concentration was reduced under fixed load forces (Figure 15, WM ensemble). When compared with other two-motor ensembles (Figure 13, load force is $0.2pN$), it is seen that at constant load force and ATP concentrations, the cargoes carried by ensembles WW, WM and MM exhibit almost the same

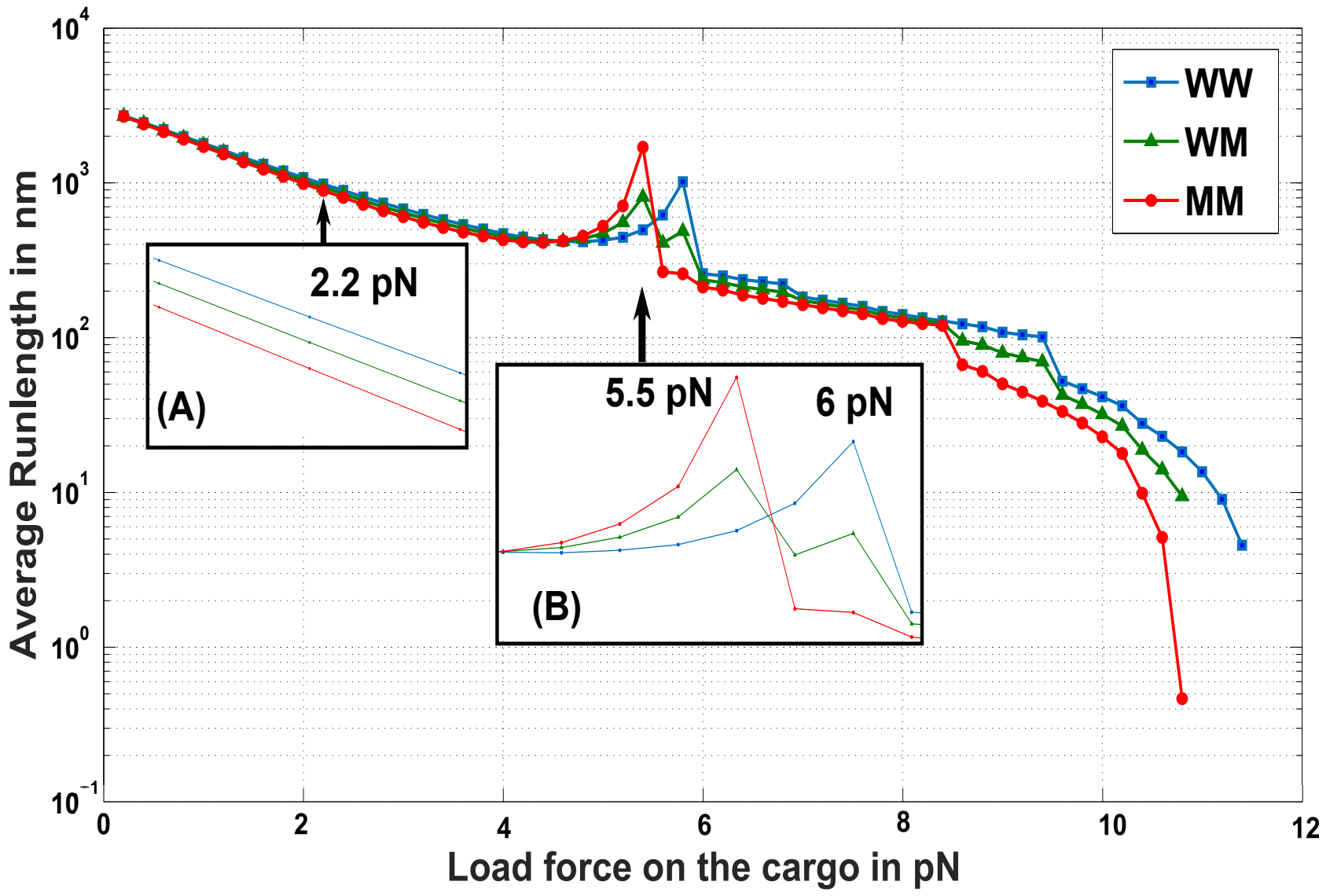

Figure 9: **Average Run-length under different load forces on a cargo with ensembles of two motors, wild type and mutant (WW, WM and MM).**

average run-length.

Thus, if the ATP concentration and number of motors within the ensemble is constant and if the load force does not approach the stall force (or multiples of stall force) for either wild type or mutant (i.e. $F_s$ or $\bar{F}_s$), then the average run-length will not be affected by the number of mutant motors in the ensemble. Similar conclusions hold for ensembles with three motors.

When load forces are close to, but less than, the stalling force $F_s$ for the same number of motors in the ensemble, then irrespective of the ATP concentration the ensembles with more wild-type motors will dominate cargo transport and exhibit longer run-lengths. In contrast, if load forces are close to, but less than $\bar{F}_s$, then ensembles with more mutant motors will dominate cargo transport and exhibit longer run-lengths.

# Conclusions

## Single motor mutation influences the emergent properties of cargo transport

We developed a simulation engine to study the transport properties of a cargo carried by multiple motor proteins of different types. To demonstrate the efficacy of our methodology we capitalized on a previous study by Selvin and coworkers [16] which characterized a kinesin motor that exhibits a reduced stall force. The study reported that placing a negative charge at the Ser-175 position of the kinesin motor domain, through either mutation or phosphorylation, reduces the stalling force of the motor and decreases its average velocity under external loads. We incorporate this into our method to enable an analysis of the behavior of ensembles comprising of several wild-type and mutant kinesin motors, where the mutant motors have lower stalling forces. The analysis provides support for the

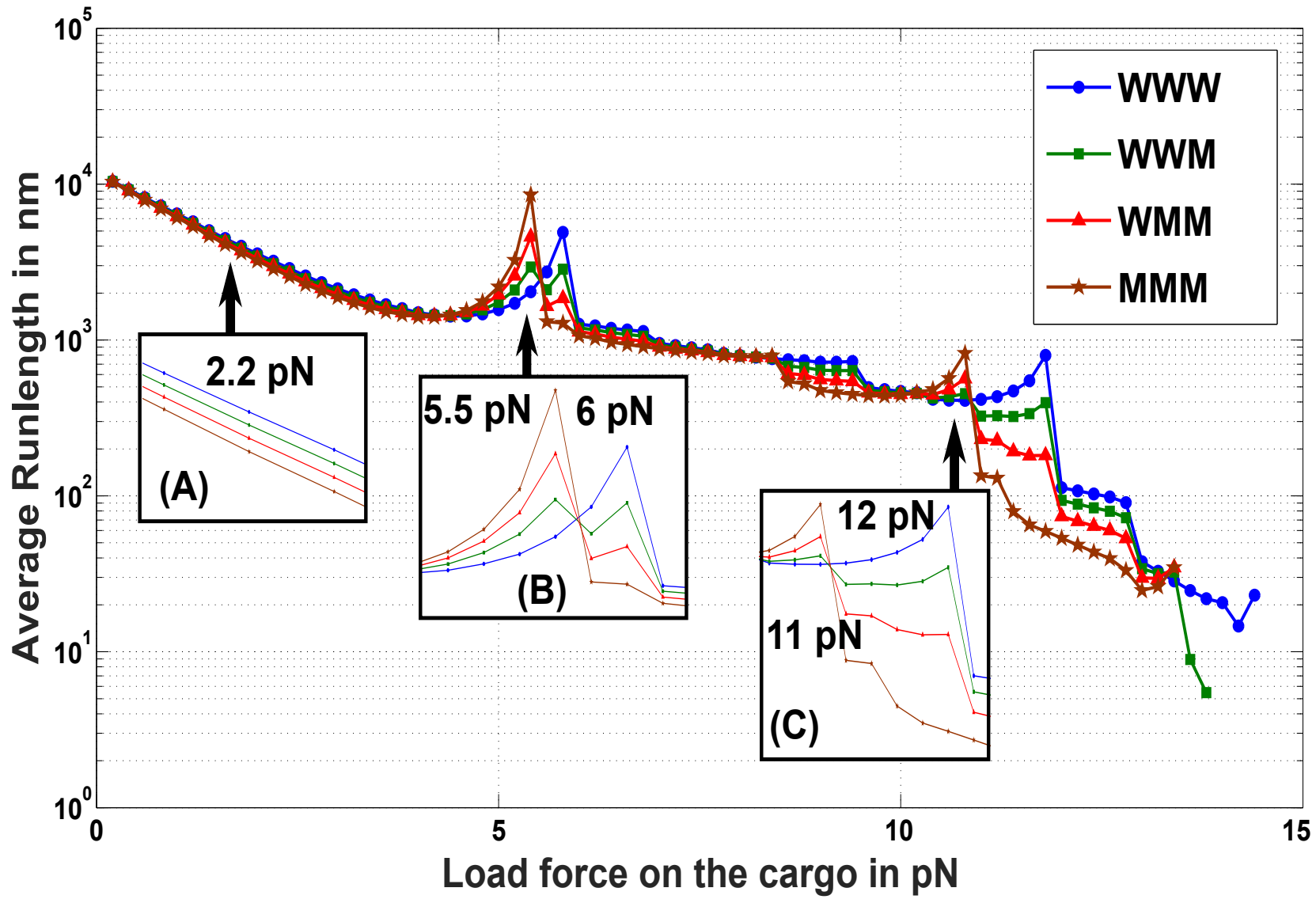

Figure 10: **Average Run-length under different load forces on a cargo with ensembles of three motors, wild type and mutant (WWW, WWM, WMM and MMM).**

following conclusions :

1. A mutation that modulates stalling force of the motors impacts the emergent transport properties of a molecular motor ensemble as follows :

   (a) For a cargo bound to a fixed number of motors and under a constant load force away from the mutant motor stalling force $\bar{F}_s$ (or its multiples), the average velocity and run-length decreases with an increase in the number of mutant motors in the ensemble. A similar trend is observed for a load force close to the wild-type motor's stalling force $F_s$ (or its multiples).

   Such a behavior is expected, since cargoes carried by ensembles having a higher proportion of mutant motors (that have a reduced stalling force) would traverse shorter run-lengths on average. Under these conditions, ensembles with a higher mutant population will eventually lag behind similar-sized ensembles with a lower mutant population.

   (b) Remarkably, for a cargo subjected to load forces close to the mutant motor's stalling force $\bar{F}_s$ (or its multiples), the average run-length increases with an increase in the number of mutant motors in the ensemble. We hypothesize the existence of a *mooring mechanism*, where mutant motors tether the cargo to the microtubule, providing the other detached motors with a greater opportunity to reattach to the mirotubule and move the cargo forward. It consequently leads to a higher average run-length of the cargo.

   Thus, surprisingly under these conditions, ensembles with a higher population of mutants gain an aggressive edge over similar sized ensembles with a lower mutant population.

2. ATP concentration modulates cargo transport by multi-motor ensembles as follows :

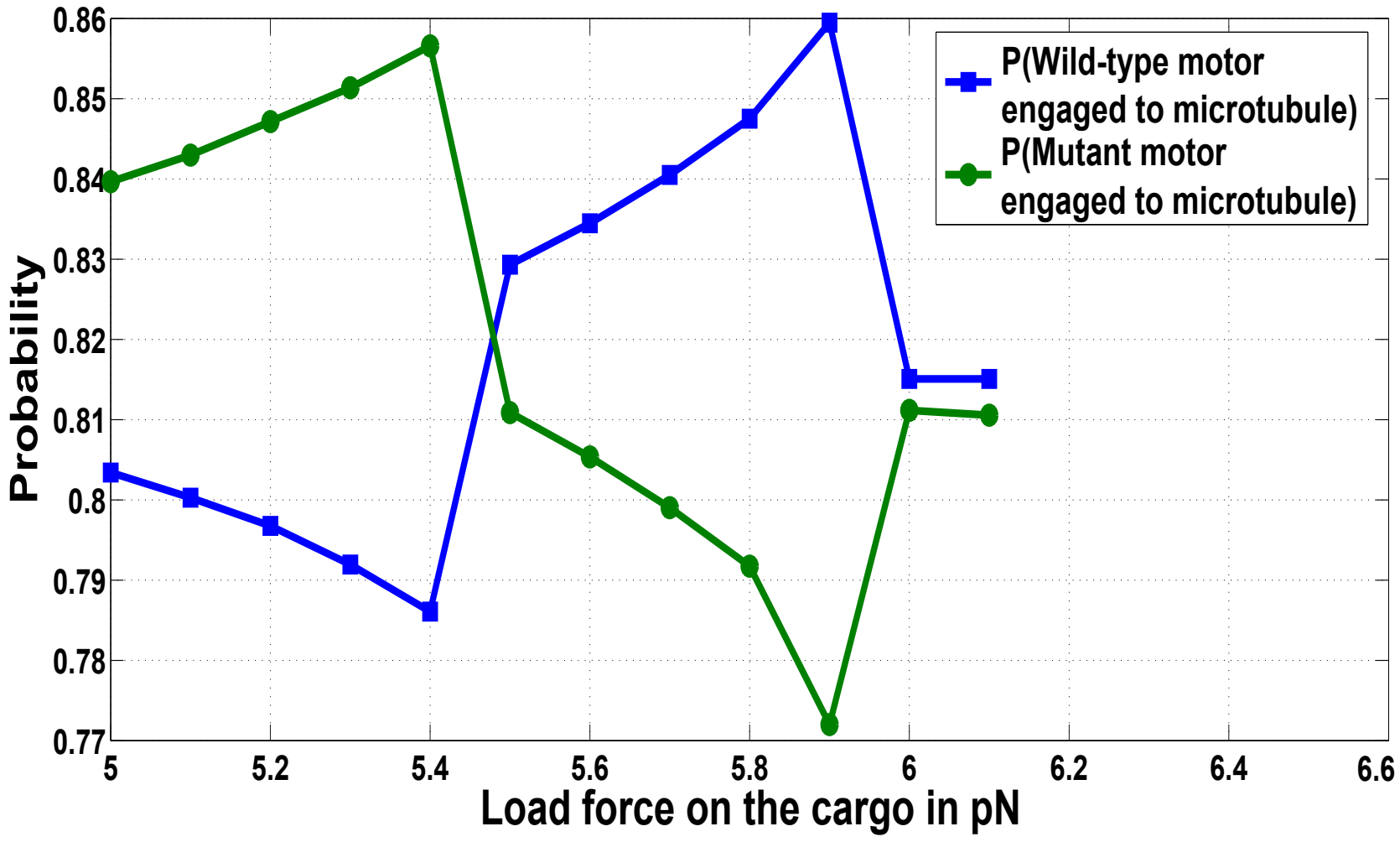

Figure 11: **Probability of wild-type and mutant motor being attached for ensemble WM.**

(a) ATP concentration in specific ranges negatively correlates with average run-length of the cargo under a constant load force, if the cargo has more than one motor attached. This finding agrees with the experimental observation made by Xu and coworkers for a 2-motor ensemble in [12].

(b) For a fixed number of motors bound to the cargo, the effect of ATP concentration on run-length is independent of the ratio of wild-type and mutant motors.

3. While we used a kinesin phosphorylation mutant [16] to explore the consequences of a multi-motor ensemble containing motors of lower stall forces, similar analysis and the computational methodology can be utilized to obtain insights on the effect of different types of motors modeled by different sets of parameters on the transport of a common cargo. Moreover, mutations that impact other single motor parameters, including on/off rates, directionality, and elasticity, could be evaluated for impacts on ensemble behavior.

It is likely that the aberrant behavior caused due to the presence of mutant motors with lower stalling forces, when occurring over multitudes of ensembles, can contribute to the disruption of cargo traffic, impediment of neuronal function and the emergence of neurodegeneration.

## Simulation engine developed for evaluating the impact of single motor parameters on behavior of motor ensembles

The computational methodology proposed by Gross and coworkers in [11] employed Monte Carlo simulations to offer several insights regarding transport properties of molecular motor ensembles. In this article, we have developed a novel, semi-analytical approach to study the effect of single motor mutations on the behavior of motor ensembles and cargo transport. This methodology allows us to compute exactly, the biologically relevant quantities such as average run-length and velocity for an ensemble of motors using experimental parameters derived from single motor experiments. The results for ensembles WW and WWW (i.e. all wild-type motors) are in exact quantitative agreement with those obtained via Monte Carlo simulations as presented in [11] as well as [13] , fully validating the methodology. Unlike Monte Carlo approaches, this model is computationally less extensive and its efficiency is independent of the number of iterations. The model also offers ways to investigate rare events and can be extended to any species of motor proteins, given the knowledge of their individual parameters. It is also easily possible to interrogate specific single motor parameters and to determine their effect on ensemble behavior, thus making it a useful tool to study the contribution of other single motor mutations as well as post translational modifications to the transport behavior of cargoes.

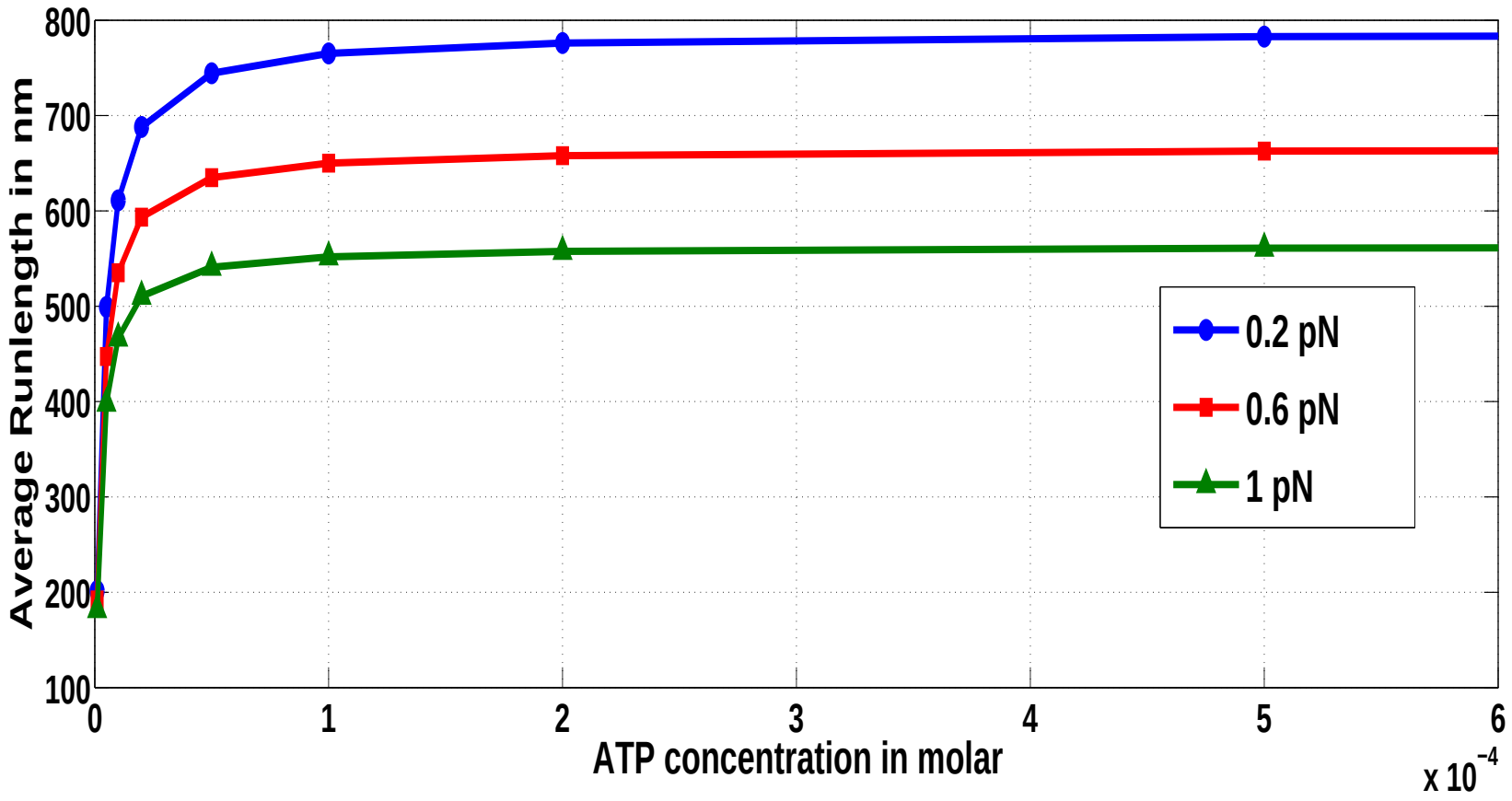

Figure 12: **Effect of ATP concentration on run-length for ensemble containing one wild-type motor, for various hindering load forces on the cargo.**

# Methods

The Master Equation based methodology used to obtain the aforementioned results is described in this section. This method is used to study emergent properties of an ensemble of multiple motors of two types, that can each take a step on, detach from or reattach onto the microtubule. The knowledge of transition probabilities of stepping, detachment and attachment enable the determination of transition rates between various *absolute configurations* of the motors, allowing for the calculation of transition rates between the corresponding *relative configurations*. These rates enable the calculation of the probability distribution of the various ensemble configurations, thereby facilitating the computation of several biologically relevant quantities such as average velocity, run-length and number of attached motors.

We begin with the construction of the relative state space, along with calculations necessary to arrive at several biologically relavent quantities. Then, a general methodology to obtain the transition rates between absolute configurations given the knowledge of the probability rates of stepping, detachment and reattachment to the microtubule for a wild-type motor$(P_S, P_D, P_A)$ and probability rates for a mutant motor$(\bar{P}_S, \bar{P}_D, \bar{P}_A)$ is presented. Finally, the model used to determine the probability rates for kinesin motor proteins is detailed.

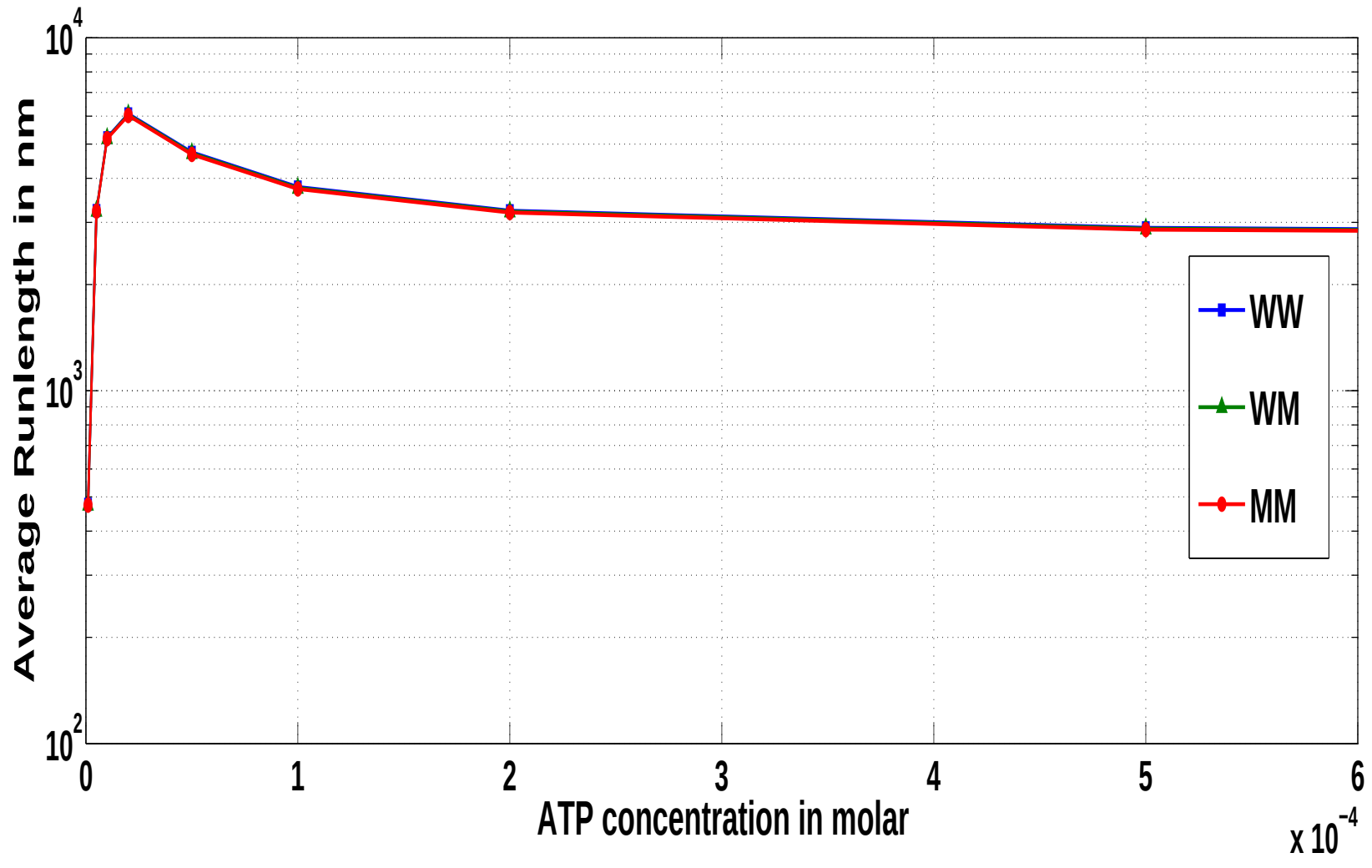

Figure 13: **Effect of ATP concentration on run-length for ensembles WW, WM and MM against load force of** $0.2$ $pN$**.**

## Construction of Relative Configuration Space

Consider a cargo that is carried by both wild-type and mutant motor proteins on a microtubule. The microtubule is modeled as directed linear lattice formed by equally sized dimers with dimension $d$. Here the $k^{th}$ dimer is located at location $\bar{a}_k = kd$ and indexed by the set of integers $I = \{..., -2, -1, 0, 1, 2, ...\}$. Each motor protein bound to the cargo can attach, take a forward step or detach from the microtubule. The *absolute configuration* $\Omega := \left\{ \begin{array}{c} \Omega_{h,k} \\ \Omega_{d,k} \end{array} \right\}_{k \in I}$ of motor-protein arrangement on the microtubule specifies the number, $\Omega_{h,k}$, of wild-type motor proteins and the number, $\Omega_{d,k}$, of mutant motor proteins at the $k^{th}$ location on the microtubule.

For example, the absolute configuration of the ensemble of motors illustrated in Figure 16 is given by

$$\Omega = [\cdots \Omega_{-1}\ \Omega_0\ \Omega_1\ \Omega_2\ \Omega_3\ \Omega_4\ \Omega_5\ \Omega_6 \cdots] = \left[ \begin{array}{ccccccccc} \cdots & 0 & 1 & 0 & 1 & 0 & 0 & 0 & 1 & \cdots \\ \cdots & 0 & 0 & 0 & 1 & 0 & 1 & 0 & 0 & \cdots \end{array} \right].$$

The *relative configuration* of an ensemble of motors is represented using a string of three symbols. Given an absolute configuration we first identify the *rearguard motor* which is the motor that is attached to the microtubule and lags behind all the other motors on the microtubule. Using the location of the reargurad motor as a reference, the relative configuration $\vartheta$ is obtained as a string of three symbols '$M_h$', '$M_d$' and '$|$', where '$M_h$' and '$M_d$' denote wild-type and mutant motors respectively, with '$|$' denoting a separator that distinguishes different microtubule locations. The motor located the farthest from the rearguard motor on the microtubule is identified as the *vanguard motor*.

For example, the relative configuration of the ensemble in Figure 17(a) is the string '$|M_h M_d||M_d||M_h|$'. The configuration that results after the furthermost mutant motor in Figure 17(a) takes a step is shown in Figure 17(b) which has a relative configuration given by '$|M_h M_d|||M_d|M_h|$'.

Both the mutant as well as the wild-type motor proteins are characterized via their own set of stepping, attachment, and detachment probabilities(for wild-type $(P_S, P_D, P_A)$ and for mutant

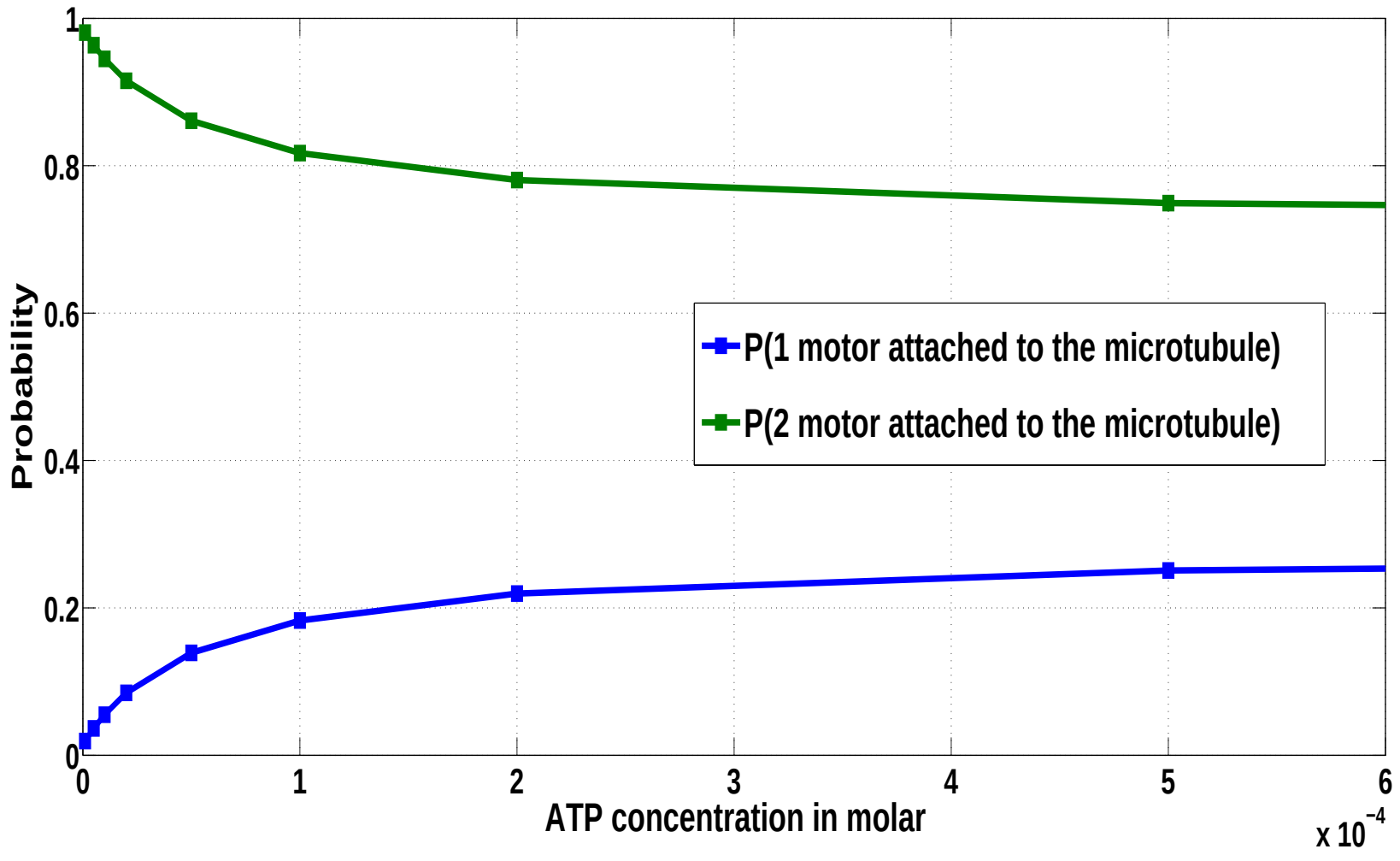

Figure 14: **Effect of ATP concentration on run-length for ensemble WW.**

$(\bar{P}_S, \bar{P}_D, \bar{P}_A))$. The individual motors for both species are modeled as hookean springs when stretched that offer no resistance when compressed. A single motor is assumed to have a linkage rest length $L_0$ and spring stiffness constant $K_e$. Motors of both the species are assumed to not step backward and are bound to the cargo particle irreversibly. More complex models of motor-proteins can be easily accommodated. It is further assumed that the there exists a force $F_{stall}$ called the *stalling force*, where if the force on the motor protein $F \geq F_{stall}$ then the motor does not take a forward step and stepping probability is zero [21]. The stalling force can be that is also used to estimate how many motors are carrying the cargo. For an ensemble of wild-type and mutant motors carrying a cargo, the following result holds:

*Result 1 : Given an ensemble of $M$ molecular motors attached to a common cargo that is subjected to a load force $F_{load}$, the distance between the rearguard and the vanguard motor is bound by*

$$n = max\left\{\frac{(M+1)max(F_s,\bar{F}_s) - F_{load}}{K_e} + d, \frac{F_{load}}{K_e}\right\} + 2L_0 \tag{1}$$

*where $F_s$ is the minimum load force for which the stepping probability of the wild-type motor protein becomes zero ( i.e. the stalling force for the wild-type motor protein), $\bar{F}_s$ is the minimum load force for which the stepping probability of the mutant motor protein becomes zero (i.e. the stalling force for the mutated motor protein), $L_0$ is the rest length of the motor linkage, $K_e$ is the linkage stiffness and $d$ is the step-size of the motor.*

A detailed derivation is provided in the Supplementary Information Text S1.

It is to be noted that the absolute configuration space admits infinitely many representations as there is always a small probability of finding the cargo at any location on the microtubule. However, the above result concludes, that given a stall force for both wild-type and mutant motors the relative configuration space is finite, since there are no motors beyond $n$ units away from the rearguard motor in any relative configuration.

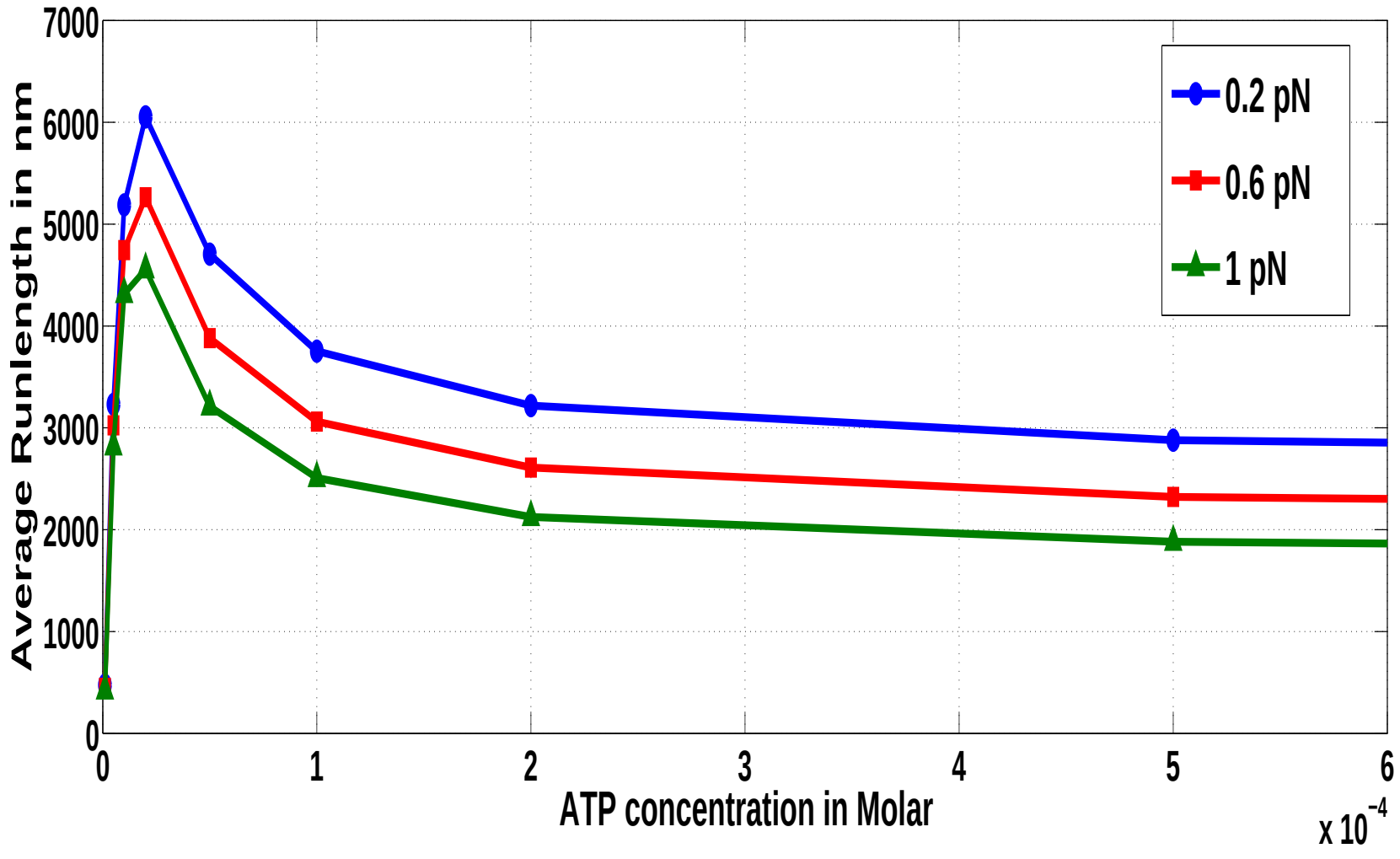

Figure 15: **Effect of ATP concentration on run-length for ensemble WM, for various hindering load forces on the cargo.**

**Transition Probabilities between Relative Configurations**

Next, we determine the probability of transitioning from a relative configuration to another from the transition probabilities in the absolute configuration space.

It is evident that an absolute configuration can be associated with a unique relative configuration. For example, the relative configuration associated with Figure 16 is $|M_h||M_hM_d||M_d||M_h|$. Thus the absolute configuration

$$\Omega = \begin{bmatrix} \cdots & 0 & 1 & 0 & 1 & 0 & 0 & 0 & 1 & \cdots \\ \cdots & 0 & 0 & 0 & 1 & 0 & 1 & 0 & 0 & \cdots \end{bmatrix}$$

is mapped to the relative configuration $|M_h||M_hM_d||M_d||M_h|$. We denote the projection operator $\Upsilon$ that maps an absolute configuration $\Omega$ to a corresponding relative configuration by $\Upsilon(\Omega)$.

The probability that the absolute configuration is $\Omega'$ at time $t+\Delta t$ conditioned on an initial configuration of $\Omega$ at $t$ is represented as $P_\Omega(\Omega', t+\Delta t|\Omega, t)$. It is assumed that the *transition probability* $P_\Omega(\Omega', \Omega)$ between the two absolute configurations $\Omega$ and $\Omega'$ is given by $\nu_\Omega(\Omega', \Omega)\Delta t$ for a small time interval $\Delta t$, where the notation $\nu_\Omega(\Omega', \Omega)$ represents the *probability rate of transition* between $\Omega$ and $\Omega'$. The underlying assumption that the rate $\nu_\Omega(\Omega', \Omega)$ is independent of the time instant $t$ holds true, since the $\nu_\Omega$ only depends upon the initial configuration and the type of transition (motor stepping, detachment or attachment) from $\Omega$ to $\Omega'$.

In a similar manner, the transition probability that the relative configuration is $\vartheta'$ at time $t+\Delta t$ given that it is $\vartheta$ at time $t$ is denoted by $P_\vartheta(\vartheta', t+\Delta t|\vartheta, t)$. The *transition probability* $P_\vartheta(\vartheta, \vartheta')$ between two relative configurations $\vartheta$ and $\vartheta'$ is expressed as $\nu_\vartheta(\vartheta', \vartheta)\Delta t$ where $\nu_\vartheta(\vartheta', \vartheta)$ denotes the probability rate of transition between $\vartheta$ and $\vartheta'$.

The knowledge of the transition rate $\nu_\Omega$ in the absolute configuration space enables the following result :

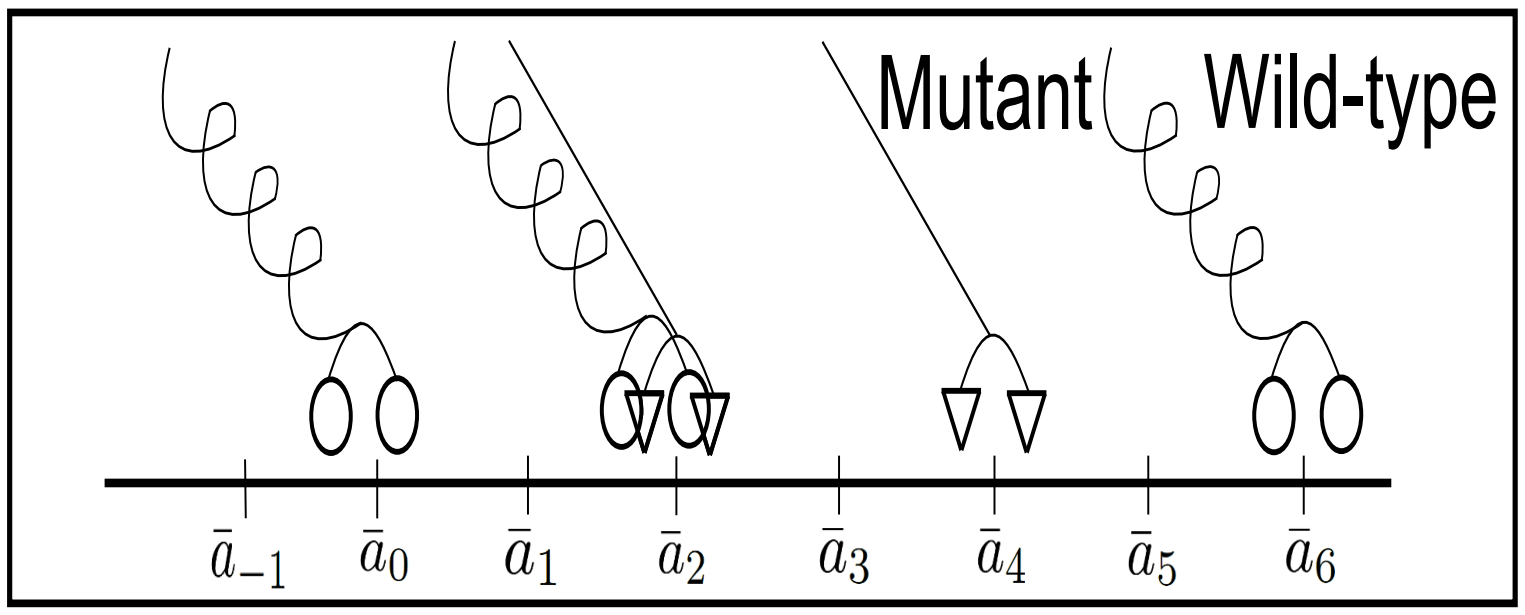

Figure 16: **Locations of wild-type and mutant motors on a section of the microtubule lattice.** The microtubule is modelled as a directed linear lattice $\bar{a}_k = \bar{a}_0 + kd$, where $\bar{a}_k$ is the position of the $k^{th}$ location. If all the motors carry a common cargo i.e. the wild-type motor at $a_0$ is the rearguard motor, the string representation is $|M_h||M_hM_d||M_d||M_h|$

*Result 2 : The rate of transition between relative configurations $\vartheta$ and $\vartheta'$ is given by :*

$$\nu_\vartheta(\vartheta', \vartheta) = \sum_{0 \leq \beta \leq \lceil \frac{n}{d} \rceil} \nu_\Omega(\tau^\beta \Omega', \Omega) \tag{2}$$

*where $\Omega$ is any absolute configuration that satisfies $\Upsilon(\Omega) = \vartheta$, $\Omega'$ is any absolute configuration that satisfies $\Upsilon(\Omega') = \vartheta'$, $(\tau^\beta \Omega')$ is an absolute configuration obtained after linearly shifting all the motors in $\Omega'$ by $\beta$ locations on the microtubule towards the right and d is the dimension of a single microtubule dimer*

A detailed derivation is provided in Supplementary Information Text S2.

As $n$ is finite for a finite number of molecular motors attached to the cargo (from (1)), () involves only a finite number of computations.

The transition rates between absolute configurations are obtainable using the chemical kinetics of the motor protein as it steps, detaches or attaches to the microtubule. The rate $\nu_\vartheta(\vartheta', \vartheta)$ is obtainable using *Result 2* and the knowledge of the rates $\nu_\Omega$ between corresponding absolute configurations.

*Results 1 and 2* together imply that given a finite number of molecular motors carrying a common cargo, the relative configuration space is finite. If $H$ is the set of all the possible relative configurations, the knowledge of transition rates from () can be used to show that the probability $P_\vartheta(\vartheta, t)$ of the relative configuration being $\vartheta$ at time $t$ satisfies the master equation,

$$\frac{\partial}{\partial t} P_\vartheta(\vartheta, t) = \sum_{\vartheta' \in H} \nu_\vartheta(\vartheta, \vartheta') P_\vartheta(\vartheta', t) - P_\vartheta(\vartheta, t) \sum_{\vartheta' \in H} \nu_\vartheta(\vartheta', \vartheta). \tag{3}$$

**Evolution of Probability Distribution of Relative Configuration**

Solution for the master equation (3) determines the time evolution of probabilities of all the relative configurations. Consider a ordering of all relative configurations given by $\vartheta_1, \ldots, \vartheta_N$ (where $N$ is the finite number of relative configurations) where the probability of finding the motors in a relative configuration $\nu_i$ at time $t$ is denoted by $P_i(t)$ and let $P(t) = [P_1(t), \ldots, P_N(t)]^T$. Using the expression of transition rates $\nu_\vartheta(\vartheta_j, \vartheta_i')$ and (3), it can be shown that the dynamics of the model describing the

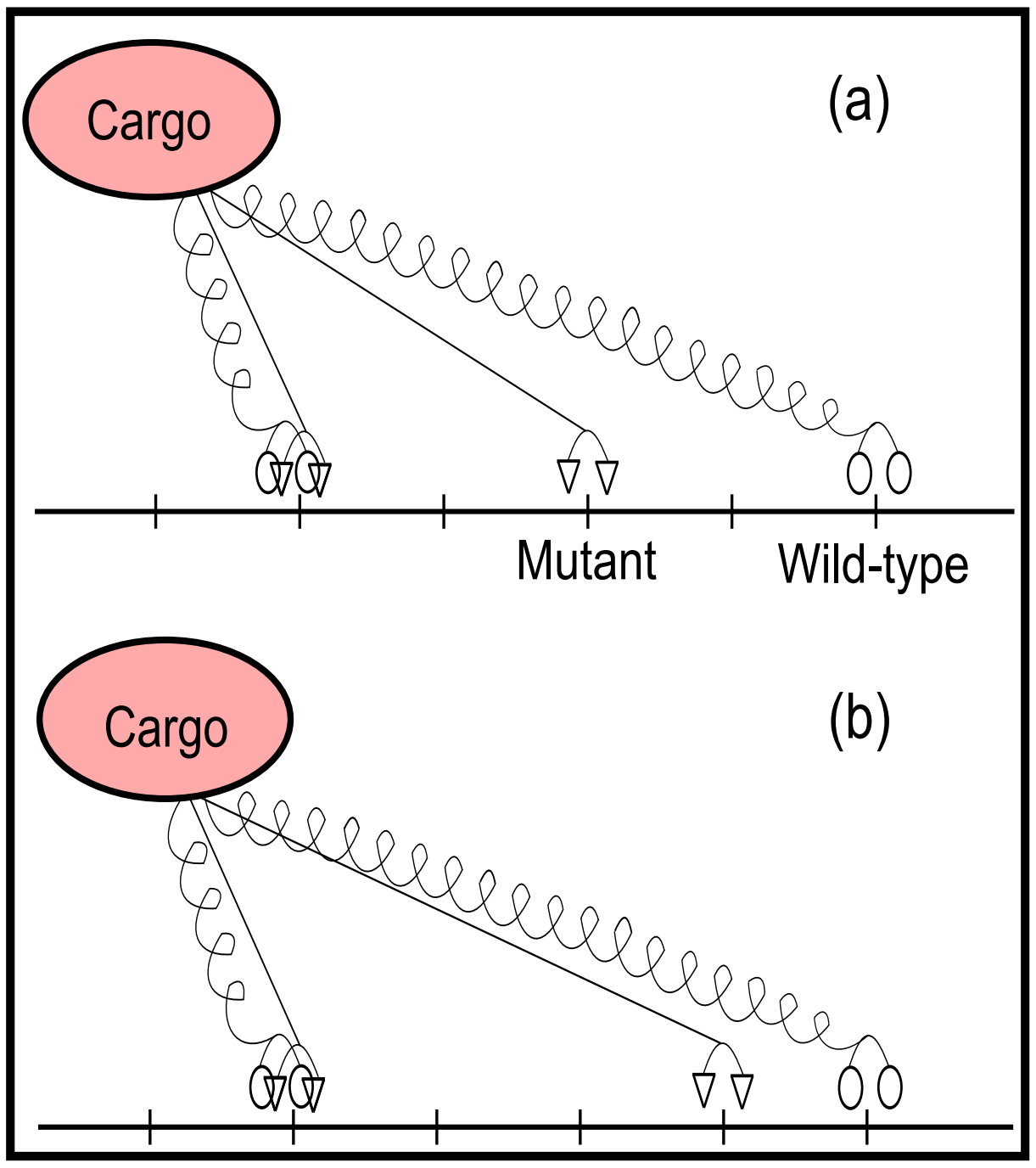

Figure 17: **Locations of wild-type and mutant motors on a section of the microtubule lattice.** The string representation for the two configurations is $(a)$ $|M_hM_d||M_d||M_h|$ and $(b)$ $|M_hM_d|||M_d|M_h|$

vector $P(t)$ is given by :

$$\frac{d}{dt}P(t) = \Gamma P(t),$$

where $\Gamma \in \mathcal{R}^{N\times N}$ is a sparse stochastic matrix which is determined by the transition rates $\nu_\vartheta(\vartheta_j, \vartheta_i)$ ($\Gamma_{ji} = \nu_\vartheta(\vartheta_j, \vartheta_i)$ if $i \neq j$, $\Gamma_{ii} = 1 - \sum_{i\neq j} \nu_\vartheta(\vartheta_j, \vartheta_i)$). Given a specified initial probability vector $P(t_0)$, it follows that

$$P(t) = e^{\Gamma(t-t_0)}P(t_0). \tag{4}$$

When specified for kinesin-1 motors and realistic values of number of motors ($M \leq 5$) and system parameters, the dimension of $\Gamma$ is $N \times N$ where $N$ lies between $10^5$ and $10^7$, making the evaluation of $e^{\Gamma(t-t_0)}$ manageable using a standard computer(the results in this article are obtained using Intel quad core i5 processor, 3.4 GHz, RAM 8 GB). The software is easily implementable using platforms like MATLAB and is faster and computationally more efficient than Monte-Carlo based approaches. Computer clusters can be used to manage more complex scenarios involving larger number of motors.

**Calculating Biologically Relevant Quantities**

Once the probability vector $P(t)$ is known, expressions of several biologically relevant quantities can be obtained in the following manner.

**Average Number of Engaged Motors (Wild-type/Mutant)**

Average number $m_h(t)$ of wild-type motors attached to the microtubule is given by

$$< m_h(t) >= \sum_{i=1}^{N} m_h(\vartheta_i) P_i(t), \tag{5}$$

where $m_h(\vartheta_i)$ represents the number of wild-type motors in the relative configuration $\vartheta_i$ or the number of $M_h$ symbols in the representation $\vartheta_i$. In a similar manner, the average number of mutant motors $m_d(t)$, attached to the microtubule is given by

$$< m_d(t) >= \sum_{i=1}^{N} m_d(\vartheta_i) P_i(t), \tag{6}$$

where $m_d(\vartheta_i)$ represents the number of mutant motors in the configuration $\vartheta_i$ or the number of $M_d$ symbols in the representation $\vartheta_i$.

**Average Velocity and Average Run-length**

Average velocity $v(t)$ of the cargo being carried by $M$ motors is determined as

$$v(t) = \sum_{\vartheta \in H} \sum_{\vartheta' \in H} d_{avg}(\vartheta', \vartheta) \nu_{\vartheta}(\vartheta', \vartheta) P_{\vartheta}(\vartheta, t), \tag{7}$$

where $d_{avg}(\vartheta', \vartheta)$ is the expected change in cargo position, when the initial and final relative configurations at $t$ and $t + \Delta t$ are restricted to being $\vartheta$ and $\vartheta'$ respectively and is given by

$$d_{avg}(\vartheta', \vartheta) = \frac{1}{\nu_{\vartheta}(\vartheta', \vartheta)} \sum_{0 \leq \beta \leq n_{max}} d(\rho^{\beta}\Omega', \Omega) \nu_{\Omega}(\rho^{\beta}\Omega', \Omega). \tag{8}$$

The expression $d(\Omega', \Omega)$ in (8) is the change in cargo equilibrium position when the absolute configuration changes from $\Omega$ to $\Omega'$. The detailed formulation of (7) and (8) is provided in Supplementary Information Text S3.

Average run-length is then calculated by summing the average velocity over time

$$Average\ Runlength = \int_{0}^{+\infty} v(t)\, dt \tag{9}$$

## Transition rates between absolute configurations

In this section we present a general scheme for determining transition rates between absolute configurations. We begin with a structural model for single motor protein that consists of motor head, stalk and cargo binding tail domain. The linkage between the motor-heads and tail for single motor is modeled as a hookean spring when stretched, that has a rest length $L_0$. It offers no resistance when compressed [11]. The motor heads move along the microtubules exerting a force $F$ on a cargo that is expressed as a function of its length $L$ by,

$$F(L) = \begin{cases} K_e(L - L_0) & \text{if } L \geq L_0, \\ 0 & \text{if } |L| < L_0, \\ K_e(L + L_0) & \text{if } L \leq -L_0. \end{cases}$$

$Z_{eq}$ is the mean position of the cargo that is its equilibrium position determined by the forces exerted by the motors on the cargo through their linkages and the load force $F_{load}$ on the cargo. The mean cargo position for a fixed $F_{load} > 0$ is a function of the absolute configuration i.e. $Z_{eq} = Z_{eq}(\Omega)$. If the cargo position is assumed to follow a truncated Gaussian distribution $\Theta(z)$ with variance $\sigma$, its probability density $\Theta(z)$ for $|z| < 3\sigma$ is given by,

$$\Theta(z) = (e^{-\frac{z^2}{2\sigma^2}})/(2\int_0^{3\sigma} e^{-\frac{z^2}{2\sigma^2}}dz).$$

The effect of thermal noise can be incorporated by determining the steady state variance $\sigma$ of the cargo position.

A transition to another configuration $\Omega'$ occurs if either the wild type or mutant motor at a location $\bar{a}_k$ steps forward to $\bar{a}_{k+1}$, detaches from the location $\bar{a}_k$ or reattaches to the location $\bar{a}_k$ on the microtubule. By representing $\Omega'$ as $\Omega + S$, $S$ is a sequence that corresponds to the type of transition(step, detach or attach) and the type of motor(wild-type or mutant) that has transitioned. The transition rate from $\Omega$ to $\Omega'$ is determined by averaging the associated probability rate over the position of the cargo.

The model of a single motor-protein is specified via the probability $P_S(F)$ of the motor taking a step, the detachment probability, $P_D(F)$, of the motor detaching from the microtubule, and the probability of attachment $P_A$ of an unattached motor-protien to the microtubule, per second. Here $F$ is the force acting on the motor which is considered positive if it is directed opposite to the motor stepping direction (e.g. kinesin *forward* stepping is towards the mictorubule + end). Here in order to calculate the transition rates between absolute configurations it is assumed that the probability rates of step, detachment and attachment are known; later we illustrate a way to compute these probabilities for kinesin motors.

**Transition Rate for Stepping**

The stepping transition of a wild-type motor from the location $\bar{a}_k$ to $\bar{a}_{k+1}$ is represented as $\Omega \xrightarrow{STEP_h} \Omega + S^{(step)}_{h,k}$, i.e.

$$\Omega = \begin{bmatrix} \cdots & \Omega_{h,k} & \Omega_{h,k+1} & \cdots \\ \cdots & \Omega_{d,k} & \Omega_{d,k+1} & \cdots \end{bmatrix} \xrightarrow{STEP_h}$$

$$\begin{bmatrix} \cdots & \Omega_{h,k} & \Omega_{h,k+1} & \cdots \\ \cdots & \Omega_{d,k} & \Omega_{d,k+1} & \cdots \end{bmatrix} + \begin{bmatrix} \cdots & -1 & +1 & \cdots \\ \cdots & 0 & 0 & \cdots \end{bmatrix}.$$

As the transition $\Omega \xrightarrow{STEP_h} \Omega + S^{(step)}_{h,k}$ occurs if any of the $\Omega_{h,k}$ wild-type motors located at $\bar{a}_k$ step forward to the position $\bar{a}_{k+1}$, the associated transition rate for stepping, $\nu_\Omega(\Omega + S^{(step)}_{h,k}, \Omega)$ is determined by averaging over the position of the cargo as,

$$\nu_\Omega(\Omega + S^{(step)}_{h,k}, \Omega) = \Omega_{h,k}\int_{Z_{eq}(\Omega)-3\sigma}^{Z_{eq}(\Omega)+3\sigma} P_S(F(z-\bar{a}_k))\Theta(z - Z_{eq}(\Omega))dz$$

where $\Omega_{h,k}$ is the number of wild-type motors located at $\bar{a}_k$ and $P_S$ is the probability of a wild type motor taking a step from the location $\bar{a}_k$ on the microtubule per second.

The stepping transition of a mutant motor from the location $\bar{a}_k$ to $\bar{a}_{k+1}$ is represented as $\Omega \xrightarrow{STEP_d} \Omega + S^{(step)}_{d,k}$, i.e.

$$\Omega = \begin{bmatrix} \cdots & \Omega_{h,k} & \Omega_{h,k+1} & \cdots \\ \cdots & \Omega_{d,k} & \Omega_{d,k+1} & \cdots \end{bmatrix} \xrightarrow{STEP_d}$$

$$\begin{bmatrix} \cdots & \Omega_{h,k} & \Omega_{h,k+1} & \cdots \\ \cdots & \Omega_{d,k} & \Omega_{d,k+1} & \cdots \end{bmatrix} + \begin{bmatrix} \cdots & 0 & 0 & \cdots \\ \cdots & -1 & +1 & \cdots \end{bmatrix}.$$

As the transition $\Omega \xrightarrow{STEP_d} \Omega + S_{d,k}^{(step)}$ occurs if any of the $\Omega_{d,k}$ mutant motors located at $\bar{a}_k$ step forward to the position $\bar{a}_{k+1}$, the associated transition rate for stepping, $\nu_\Omega(\Omega + S_{d,k}^{(step)}, \Omega)$ is determined by averaging over the position of the cargo as,

$$\nu_\Omega(\Omega + S_{d,k}^{(step)}, \Omega) = \Omega_{d,k} \int_{Z_{eq}(\Omega)-3\sigma}^{Z_{eq}(\Omega)+3\sigma} \bar{P}_S(F(z - \bar{a}_k))\Theta(z - Z_{eq}(\Omega))dz$$

where $\Omega_{d,k}$ is the number of mutant motors located at $\bar{a}_k$ and $\bar{P}_S$ is the probability of a mutant motor taking a step from the location $\bar{a}_k$ on the microtubule per second.

**Transition Rate for Attachment/Detachment**

The attachment/detachment transition of wild-type motor at location $\bar{a}_k$ is represented as $\Omega \xrightarrow{ATT_h/DET_h} \Omega \pm S_{h,k}^{(att)}$ i.e.

$$\Omega = \begin{bmatrix} \cdots & \Omega_{h,k} & \Omega_{h,k+1} & \cdots \\ \cdots & \Omega_{d,k} & \Omega_{d,k+1} & \cdots \end{bmatrix} \xrightarrow{ATT_h/DET_h}$$

$$\begin{bmatrix} \cdots & \Omega_{h,k} & \Omega_{h,k+1} & \cdots \\ \cdots & \Omega_{d,k} & \Omega_{d,k+1} & \cdots \end{bmatrix} \pm \begin{bmatrix} \cdots & 1 & 0 & \cdots \\ \cdots & 0 & 0 & \cdots \end{bmatrix},$$

where the plus sign is for attachment and minus sign is for detachment.

As the transition $\Omega \xrightarrow{ATT_h/DET_h} \Omega \pm S_{h,k}^{(att)}$ occurs if any of the $\Omega_{h,k}$ wild-type motors located at $\bar{a}_k$ detach from the microtubule, the associated transition rate for detachment, $\nu_\Omega(\Omega - S_{h,k}^{(att)}, \Omega)$ is calculated by averaging over the position of the cargo as,

$$\nu_\Omega(\Omega - S_{h,k}^{(att)}, \Omega) = \Omega_{h,k} \int_{Z_{eq}(\Omega)-3\sigma}^{Z_{eq}(\Omega)+3\sigma} P_D(F(z - \bar{a}_k))\Theta(z - Z_{eq}(\Omega))dz$$

where $\Omega_{h,k}$ is the number of wild-type motors located at $\bar{a}_k$ and $P_D$ is the probability of a wild type motor detaching from the location $\bar{a}_k$ on the microtubule per second.

The attachment/detachment transition of mutant motor at location $\bar{a}_k$ is represented as $\Omega \xrightarrow{ATT_d/DET_d} \Omega \pm S_{d,k}^{(att)}$ i.e.

$$\Omega = \begin{bmatrix} \cdots & \Omega_{h,k} & \Omega_{h,k+1} & \cdots \\ \cdots & \Omega_{d,k} & \Omega_{d,k+1} & \cdots \end{bmatrix} \xrightarrow{ATT_d/DET_d}$$

$$\begin{bmatrix} \cdots & \Omega_{h,k} & \Omega_{h,k+1} & \cdots \\ \cdots & \Omega_{d,k} & \Omega_{d,k+1} & \cdots \end{bmatrix} \pm \begin{bmatrix} \cdots & 0 & 0 & \cdots \\ \cdots & 1 & 0 & \cdots \end{bmatrix}.$$

As the transition $\Omega \xrightarrow{ATT_d/DET_d} \Omega \pm S_{d,k}^{(att)}$ occurs if any of the $\Omega_{d,k}$ mutant motors located at $\bar{a}_k$ detach from the microtubule, the associated transition rate for detachment, $\nu_\Omega(\Omega - S_{d,k}^{(att)}, \Omega)$ is determined by averaging over the position of the cargo as,

$$\nu_\Omega(\Omega - S_{d,k}^{(att)}, \Omega) = \Omega_{d,k} \int_{Z_{eq}(\Omega)-3\sigma}^{Z_{eq}(\Omega)+3\sigma} \bar{P}_D(F(z-\bar{a}_k))\Theta(z - Z_{eq}(\Omega))dz$$

where $\Omega_{d,k}$ is the number of mutant motors located at $\bar{a}_k$ and $\bar{P}_D$ is the probability of a mutant motor detaching from the location $\bar{a}_k$ on the microtubule per second.

For both the wild-type and mutant motors, a constant probability rate of reattachment to the microtubule, $P_A$ and $\bar{P}_A$, are assumed. If a motor is linked to the cargo, it is assumed to attach to the microtubule without stretching its linkage. Thereby, the only locations where the motor reattachment can occur are located within a distance $L_0$ of the cargo.

## Probability Rates for Kinesin

The results put forth in this article correspond to an instantiation of our methodology for kinesin motor protein. The probability rates of stepping, detachment and attachment of a single kinesin motor are determined using several available studies [5, 10, 20, 21].

### Probability of Stepping, per second

Kinesin takes a step on the microtubule by hydrolyzing an ATP molecule [22] in the following manner :

$$\mathcal{M} + ATP \underset{k_{off}}{\overset{k_{on}}{\rightleftharpoons}} \mathcal{M}\ ATP \xrightarrow{k_{cat}} \mathcal{M} + ADP + p_i + \Delta E,$$

where $\Delta E$ is the energy released. From [23], the ATP hydrolysis rate predicted using Michaelis-Menten dynamics relates to the probability rate stepping for a wild-type kinesin by assuming that the free motor head binds to the microtubule location with defined probability $\eta$. $P_S$ is then expressed as

$$P_S = \frac{k_{cat}[ATP]}{[ATP] + k_m}\eta, \tag{10}$$

where $k_m = \frac{k_{cat}+k_{off}}{k_{on}}$.

From [11], the force $F$ exerted by the cargo on the motor affects motor dynamics by modifying $\eta$ as,

$$\eta(F) = \begin{cases} 1 & \text{if } F = 0, \\ 1 - (\frac{F}{F_s})^2 & \text{if } 0 < F < F_s, \\ 0 & \text{otherwise.} \end{cases} \tag{11}$$

Furthermore [11] assumes that the force $F$ influences the kinetics of ATP hydrolysis by affecting $k_{off}$ in the following manner :

$$k_{off}(F) = k_{0,off} e^{\frac{F\delta_l}{K_b T}},$$

where $k_{0,off}$ is the backward reaction rate of hydrolysis when $F = 0$, $\delta_l$ is an experimentally determinable parameter, $T$ is the temperature and $K_b$ is the Boltzmann constant. Thus, the probability rate of stepping for a wild-type motor under a constant force $F$ is given by,

$$P_S(F) = \frac{k_{cat}[ATP]}{[ATP] + \frac{k_{cat}+k_{off}(F)}{k_{on}}}\eta(f). \tag{12}$$

In the case of a mutant motor protein with an altered stalling force $\bar{F}_s$, the probability rate of stepping under a force $F$ is given by,

$$\bar{P}_S(F) = \frac{k_{cat}[ATP]}{[ATP] + \frac{k_{cat}+k_{off}(F)}{k_{on}}}\bar{\eta}(F), \tag{13}$$

where the probability $\bar{\eta}$ that the free mutant motor head binds to the microtubule location is written as

$$\bar{\eta}(F) = \begin{cases} 1 & \text{if } F = 0 \\ 1 - (\frac{F}{\bar{F}_s})^2 & \text{if } 0 < F < \bar{F}_s \\ 0 & \text{otherwise} \end{cases} \tag{14}$$

**Probability of Detachment, per second**

Kinesin motors are processive species that take a certain number of steps during their ATP driven movement along microtubule before dissociating from it. The *processivity* (denoted by $\mathcal{L}$) represents how far the motor can move, on average, before its detachment from microtubule. From definition of $\mathcal{L}$ in [24],

$$\mathcal{L} = \frac{d[ATP]Ae^{-F\delta_l K_b T}}{[ATP] + B(1+A)e^{-F\delta_l K_b T}},$$

where $A$,$B$ and $\delta_l$ are experimentally determinable parameters. The relation between the probability rate of stepping $P_S$ and detachment $P_D$ is given by,

$$\frac{P_S(F)}{P_D(F)} = \frac{\mathcal{L}}{d}. \tag{15}$$

As long as the wild-type motor is not stalled i.e. $F < F_s$, the probability rate of detachment is given by,

$$P_D(F) = \frac{[ATP] + B(1+A)e^{-F\delta_l K_b T}}{[ATP]Ae^{-F\delta_l K_b T}} P_S(F). \tag{16}$$

When the load force on the wild-type motor equals to or exceeds its stalling force i.e. for $F \geq F_s$, a constant detachment rate is assumed in [11] as,

$$P_D(F) = P_{back} = 2/sec.$$

For a mutant motor that is not stalled i.e. $F < \bar{F}_s$, the probability rate of detachment is given by,

$$\bar{P}_D(F) = \frac{[ATP] + B(1+A)e^{-F\delta_l K_b T}}{[ATP]Ae^{-F\delta_l K_b T}} \bar{P}_S(F). \tag{17}$$

When the load force on the mutant motor equals to or exceeds its stalling force i.e. for $F \geq \bar{F}_s$, the constant detachment rate is assumed to be

$$\bar{P}_D(F) = P_{back} = 2/sec.$$

**Probability of Attachment, per second**

Probability of attachment is experimentally found to be $P_A \approx 5/sec$ [25, 26]. In this article it is assumed that the probability of attachment for a mutant motor, $\bar{P}_A \approx 5/sec$.

**Numerical parameters for Kinesin**
The numerical parameters considered for wild-type motors in this article are the same as that used in [12], all of which have been experimentally verified. Specifically $k_{on} = 2.10^6\ M^{-1}s^{-1}$, $k_{off} = 55\ s^{-1}$, $k_{cat} = 105\ s^{-1}$, $F_s = 6\ pN$, $d = 8\ nm$, $d_l = 1.6\ nm$, $\delta_l = 1.3\ nm$, $A = 107$, $B = 0.029\ \mu M$, $K_{el} = 0.32\ pN/nm$ and $T = 300K$. All the parameters for mutant motors are the same, except for its stalling force which is taken as $\bar{F}_s = 5.5\ pN$.

# Acknowledgments

The authors declare no competing interests. The authors acknowledge the support of University of Minnesota Supercomputing Institute (MSI) for lending their resources.

# Supporting Information

## Supplementary Information Text S1

Here, the validity of Result 1 is established.
*Result 1 : Given an ensemble of $M$ molecular motors attached to a common cargo that is subjected to a load force $F_{load}$, the distance between the rearguard and the vanguard motor is bound by*

$$n = max\left\{\frac{(M+1)max(F_s, \bar{F}_s) - F_{load}}{K_e} + d, \frac{F_{load}}{K_e}\right\} + 2L_0$$

*where $F_s$ is the minimum load force for which the stepping probability of the wild-type motor protein becomes zero ( i.e. the stalling force for the wild-type motor protein), $\bar{F}_s$ is the minimum load force for which the stepping probability of the mutant motor protein becomes zero (i.e. the stalling force for the mutated motor protein), $L_0$ is the rest length of the motor linkage and $K_e$ is the linkage stiffness.*

Consider an ensemble of wild-type and mutant motors bound to a cargo with the following assumptions :

- There are a total of $M$ (wild-type and mutant) motors bound to the cargo. The wild-type motor stalling force is $F_s$ and the mutant motor stalling force is $\bar{F}_s$.
- The motor linkages have a rest length of $L_0$ and stiffness $K_e$. The linkages are modelled as hookean springs when stretched that offer no resistance when compressed.
- There exists a constant load force $F_{load}$ on the cargo.
- $Z_{eq}$ is the mean position of the cargo, which is the equilibrium position determined by the forces exerted on the cargo by the motors through their linkages and the load force $F_{load}$
- Any unattached motor bound to the cargo can attach to only those locations on the microtubule that are within $L_0$ distance of the cargo position.

For the discussion below, it is assumed that there is at least one motor on the cargo opposing its motion (for other cases, similar arguments can be used). Suppose there be $f$ motors assisting the cargo motion and $r$ motors opposing it ($f + r \leq M$). Let the locations of the $f$ motors on the microtubule assisting the cargo motion be $Z_1^+, Z_1^+, ..., Z_f^+$. The linkage length of a motor located at $Z_i^+$ is given by $L_i^+ = Z_i^+ - Z_{eq}$ and the force it exerts on the cargo is $F_i^+ = K_e(L_i^+ - L_0)$. If the motors are located in such a way that $0 < L_0 \leq L_1^+ \leq L_2^+ \leq ... \leq L_f^+$, then $F_1^+ \leq F_2^+ \leq ... \leq F_f^+$ and the total force exerted on the cargo in the forward direction is $F_{fwd} = \sum_{i=1}^{f} F_i^+$.

Let the locations of the $r$ motors on the microtubule opposing the cargo motion be $Z_1^-, Z_1^-, ..., Z_r^-$. The linkage length of a motor located at $Z_j^-$ is given by $L_j^- = Z_{eq} - Z_j^-$ and the force it exerts on the cargo is $F_j^- = K_e(L_j^- - L_0)$. If the motors are located in such a way that $0 < L_0 \leq L_1^- \leq L_2^- \leq ... \leq L_r^-$, then $F_1^- \leq F_2^- \leq ... \leq F_r^-$ and the total force exerted on the cargo in the backward direction is $F_{back} = \sum_{j=1}^{r} F_j^-$. The separation between the vanguard and rearguard motor can be found as $S = L_f^+ + L_r^-$ where $L_f^+$ and $L_r^-$ are the linkage lengths of the vanguard and rearguard motor respectively.

Suppose the vanguard motor (thereby all the motors) is not stalled. Then $F_{fwd} < M\ max(F_s, \bar{F}_s)$. If the vanguard motor is a wild-type motor, $F(L_f^+) < F_s$ thus $L_f^+ < \frac{F_s}{K_e} + L_0$ and if it is a mutant motor, $F(L_f^+) < \bar{F}_s$ thus $L_f^+ < \frac{\bar{F}_s}{K_e} + L_0$. In general $L_f^+ < \frac{max(F_s, \bar{F}_s)}{K_e} + L_0$. At equilibrium, $F_{back} = F_{fwd} - F_{load}$, thus $\sum_{j=1}^{r} F_j^- = F_{fwd} - F_{load}$ which implies that $F(L_r^-) + \sum_{j=1}^{r-1} F_j^- = F_{fwd} - F_{load}$.

It follows that $F(L_r^-) \leq F_{fwd} - F_{load}$. Substituting for $F(L_r^-)$ it follows that $K_e(L_r^- - L_0) < M\ max(F_s, \bar{F}_s) - F_{load}$ and thus $L_r^- < \frac{M\ max(F_s, \bar{F}_s) - F_{load}}{K_e} + L_0$. Thus the separation $S$ between the vanguard and rearguard motor when none of the motors are stalled is given as,

$$S < \frac{(M+1)max(F_s, \bar{F}_s) - F_{load}}{K_e} + 2L_0. \tag{18}$$

Let $n_{max} := \frac{(M+1)max(F_s, \bar{F}_s) - F_{load}}{K_e} + d + 2L_0$. Thus, if none of the motors are stalled then the separation $S < (n_{max} - d)$.

Consider a current configuration where the separation $S' < n_{max}$. In the current configuration, the vanguard motor is either in stalled (Case I) or not stalled (Case II) condition.

Case I : If the vanguard motor is stalled, then any ensuing change in the configuration will only decrease the separation since the vanguard motor cannot step forward and the rearguard motor cannot step backward. Thus the separation remains less than $n_{max}$.

Case II : If the vanguard motor is not stalled, then using (18) it can be stated that the separation $S' < (n_{max} - d)$. Any change in the current configuration that can increase the separation is if the vanguard motor takes a forward step of length $d$, where the new separation $(S' + d) < n_{max}$. Thus the separation remains less than $n_{max}$.

It can thus be stated that if in a current configuration the separation is less than a bound $n_{max}$ then for any subsequent configuration the bound is obeyed.

## Supplementary Information Text S2

### Transition rates between relative configurations

Here the validity of Result 2 is established.

*Result 2 : The rate of transition between relative configurations $\vartheta$ and $\vartheta'$ is given by*

$$\nu_\vartheta(\vartheta', \vartheta) = \sum_{0 \leq \beta \leq \lceil \frac{n}{d} \rceil} \nu_\Omega(\tau^\beta \Omega', \Omega),$$

*where $\Upsilon(\Omega) = \vartheta$, $\Upsilon(\Omega') = \vartheta'$, $(\tau^\beta \Omega')$ is an absolute configuration obtained after linearly shifting all the motors in $\Omega$ by $\beta$ locations on the microtubule towards the right and d is the periodicity of the microtubule lattice*

Let $\Omega$ and $\Omega'$ be a pair of absolute configurations such that $\Upsilon(\Omega) = \vartheta$ and $\Upsilon(\Omega') = \vartheta'$. It is assumed that the system is translation invariant with its stochastic behavior unaffected if the cargo and the ensemble of motors bound to it shift to a new location along the microtubule. The translation invariance property enables the construction of a set $\Omega(\vartheta)$ of all absolute configurations that have the same relative configuration $\vartheta$ as $\Omega(\vartheta) = \{\tau^{\beta}\Omega : \beta \in I\}$, where $I$ is the set of integers. In a similar manner, $\Omega(\vartheta') = \{\tau^{\beta'}\Omega' : \beta' \in I\}$

Using the definition of transition probability, $P_{\vartheta}(\vartheta', t+\Delta t|\vartheta, t)$ can be expressed as follows :

$$
\begin{aligned}
P_{\vartheta}(\vartheta', t+\Delta t|\vartheta, t) &= \frac{P_{\vartheta}(\vartheta', t+\Delta t, \vartheta, t)}{P_{\vartheta}(\vartheta, t)} \\
&= \frac{\displaystyle\sum_{\Omega\in\Omega(\vartheta)}\sum_{\Omega'\in\Omega(\vartheta')} P_{\Omega}(\Omega', t+\Delta t, \Omega, t)}{\sum_{\Omega\in\Omega(\vartheta)} P_{\Omega}(\Omega, t)} \\
&= \frac{1}{\sum_{\beta} P_{\Omega}(\tau^{\beta}\Omega, t)} \quad \sum_{\beta}\sum_{\beta'} P_{\Omega}(\tau^{\beta'}\Omega', t+\Delta t, \tau^{\beta}\Omega, t) \\
&= \frac{1}{\sum_{\beta} P_{\Omega}(\tau^{\beta}\Omega, t)} \quad \sum_{\beta}\sum_{\beta'} P_{\Omega}(\tau^{\beta'}\Omega', t+\Delta t|\tau^{\beta}\Omega, t) P_{\Omega}(\tau^{\beta}\Omega, t) \\
&= \frac{1}{\sum_{\beta} P_{\Omega}(\tau^{\beta}\Omega, t)} \sum_{\beta} P_{\Omega}(\tau^{\beta}\Omega, t) \sum_{\beta'} P_{\Omega}(\tau^{\beta'}\Omega', t+\Delta t|\tau^{\beta}\Omega, t) \\
&= \frac{1}{\sum_{\beta} P_{\Omega}(\tau^{\beta}\Omega, t)} \sum_{\beta} P_{\Omega}(\tau^{\beta}\Omega, t) \sum_{\beta'} P_{\Omega}(\tau^{(\beta'-\beta)}\Omega', t+\Delta t|\Omega, t) \\
&= \frac{1}{\sum_{\beta} P_{\Omega}(\tau^{\beta}\Omega, t)} \sum_{\beta} P_{\Omega}(\tau^{\beta}\Omega, t) \sum_{\beta'} P_{\Omega}(\tau^{\beta'}\Omega', t+\Delta t|\Omega, t) \\
&= \sum_{\beta'} P_{\Omega}(\tau^{\beta'}\Omega', t+\Delta t|\Omega, t) \\
&= \sum_{\beta'} \nu_{\Omega}(\tau^{\beta'}\Omega', \Omega)\Delta t.
\end{aligned}
$$

In the third equality, $\sum_{\beta} P_{\Omega}(\tau^{\beta}\Omega, t)$ is performed over all shifts $\beta$ of an absolute configuration $\Omega$ such that its projection is the relative configuration $\vartheta$ i.e. $\Upsilon(\Omega) = \vartheta$. Similarly while determining $\displaystyle\sum_{\beta}\sum_{\beta'} P_{\Omega}(\tau^{\beta'}\Omega', t+\Delta t, \tau^{\beta}\Omega, t)$, the absolute configuration $\Omega'$ satisfies $\Upsilon(\Omega') = \vartheta'$.

In the sixth equality, translation invariance property is applied wherein both the absolute configurations at $t$ and $t+\Delta t$ are shifted by $\beta$ places to the left (via the operation $\tau^{-\beta}$). For the seventh equality, the set $\{\tau^{(\beta'-\beta)}\} = \{\tau^{\beta'}\}$, since $\beta'$ is any integer and $\beta$ is fixed. Moreover, since $\Omega$ has been arbitrarily chosen with the only condition that $\Upsilon(\Omega) = \vartheta$, this will hold for every $\Omega \in \Omega(\vartheta)$. Using the definition that $P_{\vartheta}(\vartheta', t+\Delta t|\vartheta, t) = \nu_{\vartheta}(\vartheta', \vartheta)\Delta t$, $\nu_{\vartheta}(\vartheta', \vartheta)\Delta t = \sum_{\beta'} \nu_{\Omega}(\tau^{\beta'}\Omega', \Omega)\Delta t$ and thus,

$$\nu_{\vartheta}(\vartheta', \vartheta) = \sum_{\beta'} \nu_{\Omega}(\tau^{\beta'}\Omega', \Omega).$$

Since for a finite number of motors bound to a cargo the maximum distance between the vanguard and rearguard motor is finite (and equal to $\lceil \frac{n}{d} \rceil$ locations on the microtubule),

$$\nu_{\vartheta}(\vartheta', \vartheta) = \sum_{0\leq\beta\leq\lceil \frac{n}{d} \rceil} \nu_{\Omega}(\tau^{\beta}\Omega', \Omega).$$

## Supplementary Information Text S3

**Average velocity of the cargo**

Here, average velocity of the cargo is determined by undertaking the following steps :

**Step I** : The expected change in cargo position when the initial and final relative configurations at $t$ and $t + \Delta t$ are restricted to being $\vartheta$ and $\vartheta'$ respectively, denoted by $d_{avg}(\vartheta', \vartheta)$, is obtained.

**Step II** : The expected change in the cargo position in time $\Delta t$, denoted as $D_{\Delta t}$ is obtained by removing the restriction of the relative configurations being $\vartheta$ and $\vartheta'$ at $t$ and $t + \Delta t$. The average velocity of the cargo at time $t$ follows as,

$$v(t) = \frac{D_{\Delta t}}{\Delta t}$$

To obtain $d_{avg}(\vartheta', \vartheta)$ the following steps are undertaken :

1. The change in cargo equilibrium position,

$$d(\Omega', \Omega) = Z_{eq}(\Omega') - Z_{eq}(\Omega),$$

is determined for every possible transition from an initial absolute configuration $\Omega$ at time $t$ to a final absolute configuration $\Omega'$ at time $t + \Delta t$, where $\Omega$ and $\Omega'$ are chosen such that $\Upsilon(\Omega) = \vartheta$ and $\Upsilon(\Omega') = \vartheta'$.

2. The probability $P(\Omega', t + \Delta t, \Omega, t | \vartheta', t + \Delta t, \vartheta, t)$ of transitioning from $\Omega$ to $\Omega'$ in time $\Delta t$ is determined for every such pair of absolute configurations $(\Omega, \Omega')$, conditioned on the fact that the relative configuration also transitions from $\vartheta$ to $\vartheta'$ in the same time $\Delta t$.

3. A weighted sum of $d(\Omega', \Omega)$ with the weights given by the probabilities $P(\Omega' t + \Delta t, \Omega, t | \vartheta', t + \Delta t, \vartheta, t)$ is obtained.

Starting with any pair of absolute configurations $\Omega$ and $\Omega'$ satisfying the conditions $\Upsilon^{(e)}(\Omega) = \vartheta$ and $\Upsilon^{(e)}(\Omega') = \vartheta'$, the expected change in cargo position $d_{avg}(\vartheta', \vartheta)$, when the initial and final relative configurations at $t$ and $t + \Delta t$ are restricted to being $\vartheta$ and $\vartheta'$ respectively, is given by

$$
\begin{aligned}
d_{avg}(\vartheta', \vartheta) :&= \sum_{\Omega \in \Omega(\vartheta)} \sum_{\Omega' \in \Omega(\vartheta')} d(\Omega', \Omega) P(\Omega', t+\Delta t, \Omega, t | \vartheta', t+\Delta t, \vartheta, t) \\
&= \sum_{\beta} \sum_{\beta'} d(\tau^{\beta'}\Omega', \tau^{\beta}\Omega) P(\tau^{\beta'}\Omega', t+\Delta t, \tau^{\beta}\Omega, t | \vartheta', t+\Delta t, \vartheta, t) \\
&= \sum_{\beta} \sum_{\beta'} d(\tau^{\beta'}\Omega', \tau^{\beta}\Omega) \ \frac{P(\tau^{\beta'}\Omega', t+\Delta t, \tau^{\beta}\Omega, t, \vartheta', t+\Delta t, \vartheta, t)}{P(\vartheta', t+\Delta t, \vartheta, t)} \\
&= \sum_{\beta} \sum_{\beta'} d(\tau^{\beta'}\Omega', \tau^{\beta}\Omega) \ \frac{P(\tau^{\beta'}\Omega', t+\Delta t, \tau^{\beta}\Omega, t)}{P(\vartheta', t+\Delta t, \vartheta, t)} \\
&= \sum_{\beta} \sum_{\beta'} d(\tau^{\beta'}\Omega', \tau^{\beta}\Omega) \ \frac{\nu_{\Omega}(\tau^{\beta'}\Omega', \tau^{\beta}\Omega) P_{\Omega}(\tau^{\beta}\Omega, t)}{\nu_{\vartheta}(\vartheta', \vartheta) P_{\vartheta}(\vartheta, t)} \\
&= \frac{1}{\nu_{\vartheta}(\vartheta', \vartheta) P_{\vartheta}(\vartheta, t)} \sum_{\beta} P_{\Omega}(\tau^{\beta}\Omega, t) \ \sum_{\beta'} d(\tau^{\beta'}\Omega', \tau^{\beta}\Omega) \nu_{\Omega}(\tau^{\beta'}\Omega', \tau^{\beta}\Omega) \\
&= \frac{1}{\nu_{\vartheta}(\vartheta', \vartheta) P_{\vartheta}(\vartheta, t)} \sum_{\beta} P_{\Omega}(\tau^{\beta}\Omega, t) \ \sum_{\beta'} d(\tau^{(\beta'-\beta)}\Omega', \Omega) \nu_{\Omega}(\tau^{(\beta'-\beta)}\Omega', \Omega) \\
&= \frac{1}{\nu_{\vartheta}(\vartheta', \vartheta) P_{\vartheta}(\vartheta, t)} \sum_{\beta} P_{\Omega}(\tau^{\beta}\Omega, t) \ \sum_{\beta'} d(\tau^{\beta'}\Omega', \Omega) \nu_{\Omega}(\tau^{\beta'}\Omega', \Omega) \\
&= \frac{1}{\nu_{\vartheta}(\vartheta', \vartheta)} \sum_{\beta'} d(\tau^{\beta'}\Omega', \Omega) \nu_{\Omega}(\tau^{\beta'}\Omega', \Omega).
\end{aligned}
$$

In the seventh equality, translation invariance property is applied wherein both the absolute configurations at $t$ and $t + \Delta t$ are shifted by $\beta$ places to the left ( via. the operation $\tau^{-\beta}$). For the eighth equality, the set $\{\tau^{(\beta'-\beta)}\} = \{\tau^{\beta'}\}$, since $\beta'$ is any integer and $\beta$ is fixed. The result is identical for any choice of absolute configuration $\Omega$ that satisfies $\Upsilon^{(e)}(\Omega) = \vartheta$. Thus,

$$
d_{avg}(\vartheta', \vartheta) = \frac{1}{\nu_{\vartheta}(\vartheta', \vartheta)} \sum_{\beta'} d(\tau^{\beta'}\Omega', \Omega) \nu_{\Omega}(\tau^{\beta'}\Omega', \Omega).
$$

After obtaining the expected value $d_{avg}(\vartheta', \vartheta)$, the expected change $D_{\Delta t}$ in cargo position in time $\Delta t$ is determined to be,

$$
\begin{aligned}
D_{\Delta t} &= \sum_{\vartheta \in H} \sum_{\vartheta' \in H} d_{avg}(\vartheta', \vartheta) P_{\vartheta}(\vartheta', t+\Delta t, \vartheta, t), \\
&= \sum_{\vartheta \in H} \sum_{\vartheta' \in H} d_{avg}(\vartheta', \vartheta) \nu_{\vartheta}(\vartheta', \vartheta) P_{\vartheta}(\vartheta, t)(\Delta t).
\end{aligned}
$$

where $H$ is the set of all possible relative configurations. It enables the calculation of average velocity as,

$$
\begin{aligned}
v(t) &= \frac{D_{\Delta t}}{\Delta t} \\
&= \sum_{\vartheta \in H} \sum_{\vartheta' \in H} d_{avg}(\vartheta', \vartheta) \nu_{\vartheta}(\vartheta', \vartheta) P_{\vartheta}(\vartheta, t).
\end{aligned}
$$

## Supplementary Figures

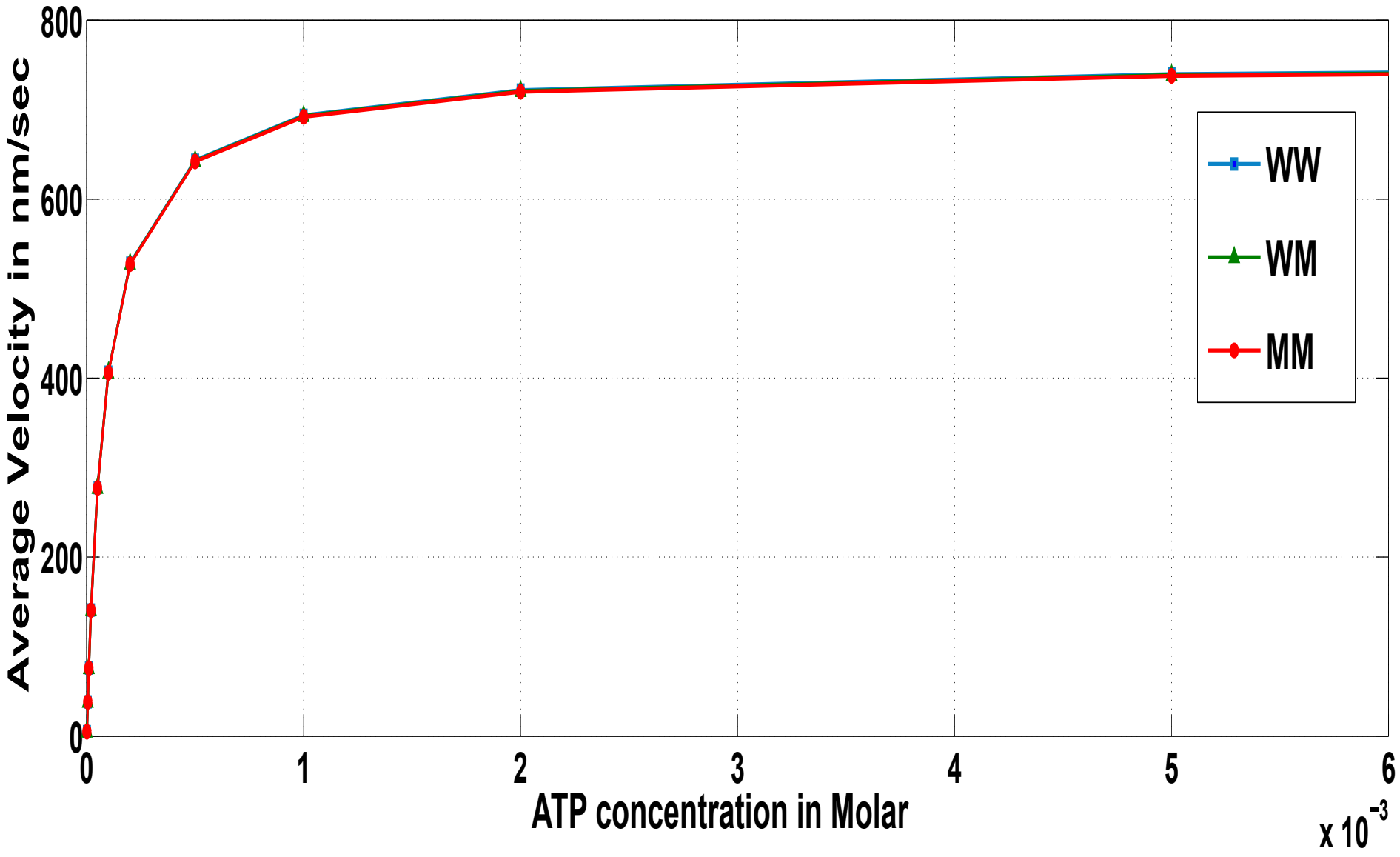

Figure 18: (**S1Fig**) Effect of ATP concentration on average velocity for 2-motor ensembles WW, WM and MM against load force of 0.2 $pN$

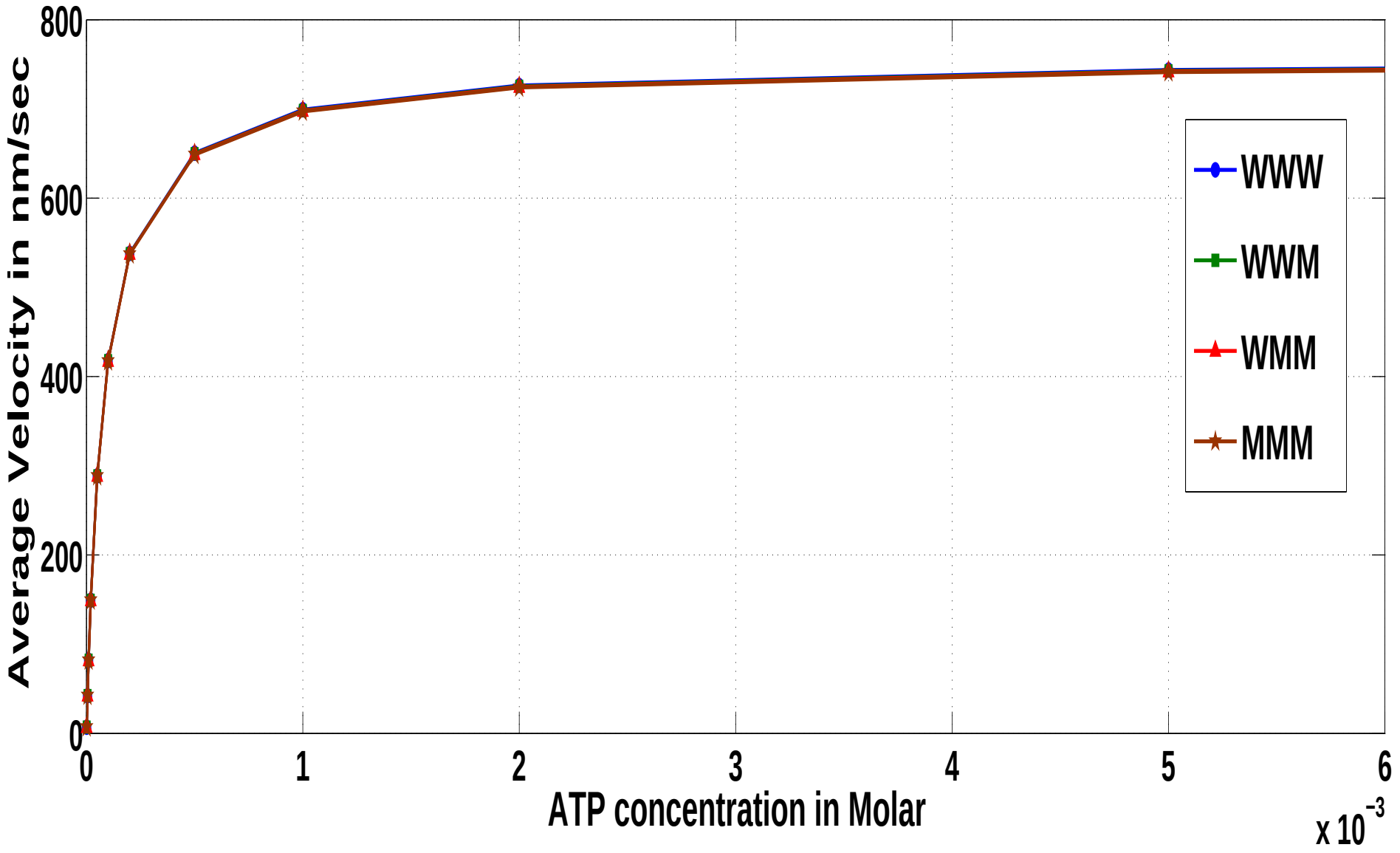

Figure 19: (**S2Fig**) Effect of ATP concentration on average velocity for 3-motor ensembles WWW, WWM, WMM and MMM against load force of 0.2 $pN$

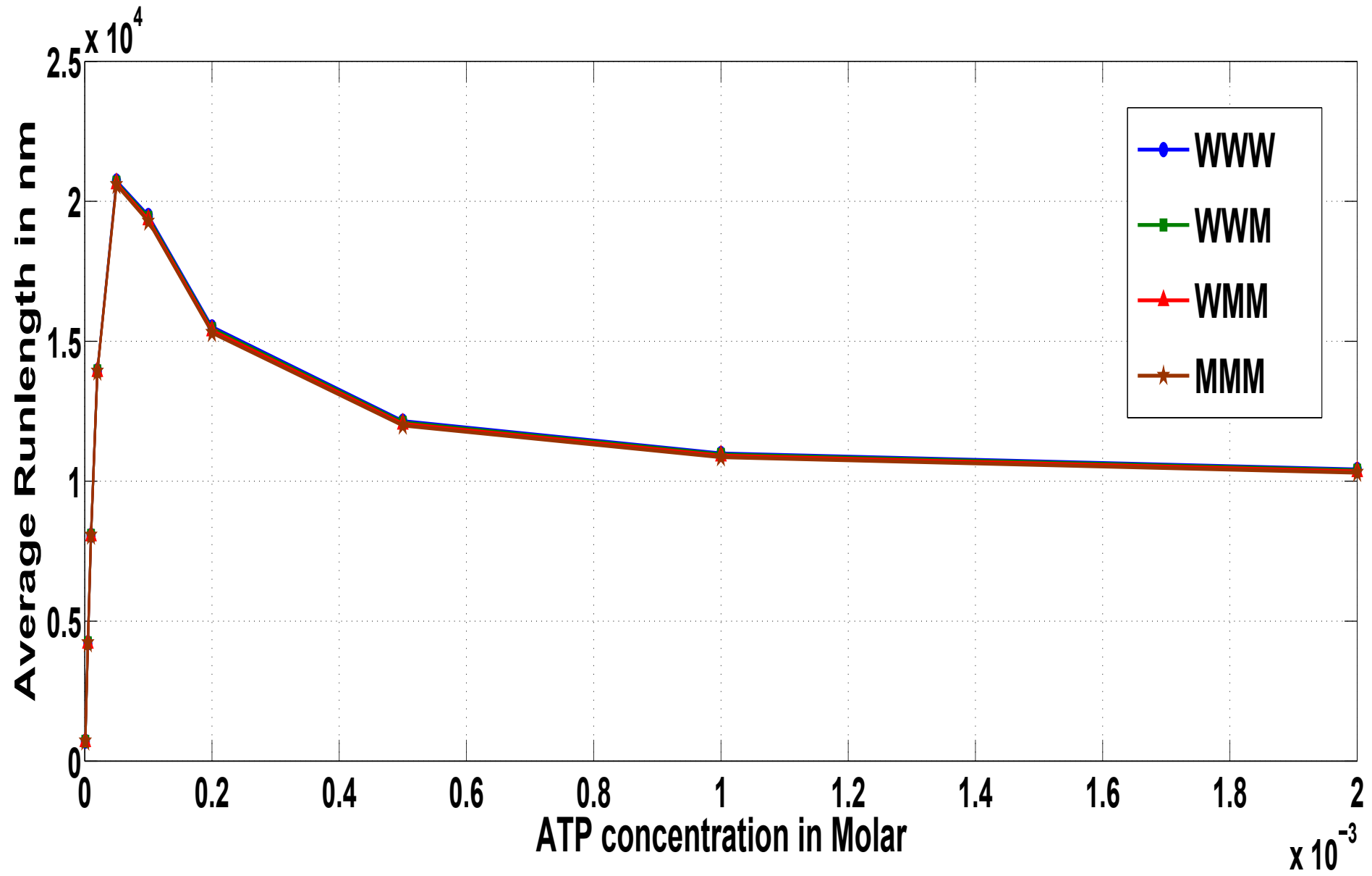

Figure 20: (**S3Fig**) Effect of ATP concentration on average runlength for 3-motor ensembles WWW, WWM, WMM and MMM against load force of 0.2 $pN$